\documentclass[aps,prd,showpacs,twocolumn, groupedaddress,nofootinbib]{revtex4}

\usepackage{amsmath, amsfonts, graphicx}

\newcommand{\Long}{\mathbb{L}}
\newcommand{\LapLong}{\Delta_\Long}
\newcommand{\tildeLapLong}{\tilde\Delta_\Long}
\newcommand{\TF}{\mbox{TF}}

\begin{document}
\title{Comparing initial-data sets for binary black holes}

\author{Harald P. Pfeiffer}
\affiliation{Department of Physics, Cornell University, Ithaca, New York\ \ 14853}
\author{Gregory B. Cook}
\affiliation{Department of Physics, Wake Forest University, Winston-Salem,
         North Carolina\ \ 27109}
\author{Saul A. Teukolsky}
\affiliation{Columbia Astrophysics Laboratory, Columbia University, New York, New York 10027}
\altaffiliation[Permanent address: ]{Department of Physics, Cornell University, Ithaca, New York\ \ 14853}

\date{\today}

\begin{abstract}
  We compare the results of constructing binary black hole initial
  data with three different decompositions of the constraint equations
  of general relativity.  For each decomposition we compute the
  initial data using a superposition of two Kerr-Schild black holes to
  fix the freely specifiable data.  We find that these initial-data sets
  differ significantly, with the ADM energy varying by as much as 5\%
  of the total mass.  We find that all initial-data sets currently
  used for evolutions might contain unphysical gravitational radiation
  of the order of several percent of the total mass.  This is
  comparable to the amount of gravitational-wave energy observed
  during the evolved collision.  More astrophysically realistic
  initial data will require more careful choices of the freely
  specifiable data and boundary conditions for both the metric and
  extrinsic curvature.  However, we find that the choice of extrinsic
  curvature affects the resulting data sets more strongly than the
  choice of conformal metric.
\end{abstract}
 \maketitle

\section{Introduction}

Numerical evolutions of black holes have been improved slowly but
steadily over the last few years and now first attempts are being made
to extract physical information from these evolutions. Most notably
one wants to predict the gravitational radiation emitted during black
hole coalescence \cite{Baker-Bruegmann-etal:2001,Alcubierre:2001,
Baker-Campanelli-etal:astroph0202469}.

The quality of the initial data will be crucial to the success of the
predictions of the gravitational wave forms. Unphysical gravitational
radiation present in the initial data will contribute to the
gravitational waves computed in an evolution and might overwhelm the
true gravitational wave signature of the physical process under
consideration.  Therefore an important question is how to control the
gravitational wave content of initial-data sets, and how to specify
{\em astrophysically} relevant initial data with the appropriate
gravitational wave content, for e.g.\ two black holes orbiting each
other.  Unfortunately, assessing and controlling the gravitational wave
content of initial-data sets is not well understood at all.

The mere {\em construction} of an initial-data set alone is
fairly involved, since every initial-data set must satisfy a rather
complicated set of four partial differential equations, the so-called
constraint equations of general relativity.  The question of how to
solve these equations, and how to specify initial data representing
binary black holes in particular, has received considerable attention.

We consider in this paper three different approaches that transform
the constraint equations into elliptic equations: The
{\em conformal transverse-traceless (TT) decomposition}\cite{York:1979},
the {\em physical TT decomposition} \cite{ Murchadha-York:1974a,
  Murchadha-York:1974b, Murchadha-York:1976} and the {\em conformal
  thin sandwich decomposition}\cite{York:1999}.  These decompositions
split the variables on the initial-data surface into various pieces in
such a way that the constraint equations determine some of the pieces while
not restricting the others.  After these freely
specifiable pieces are chosen, the constraint equations are solved and the
results are combined with the freely specifiable pieces to yield a
valid initial-data set.

Any reasonable choice for the freely specifiable pieces will lead to a
valid initial-data set. Furthermore, any one of these decompositions
can generate any desired initial-data set, given the {\em correct}
choices of the freely specifiable pieces.  However, it is not clear
{\em what} choices of freely specifiable pieces lead to initial-data
sets with the desired properties.

The decompositions we consider here lead to four coupled nonlinear
elliptic partial differential equations. Since such equations are difficult to
solve, the early approach to constructing initial data was pragmatic:
One used the conformal TT decomposition with additional restrictions
on the freely specifiable pieces, most notably conformal flatness and
maximal slicing. These assumptions decouple the constraints
and allow for analytical solutions to the momentum
constraints, the so-called {\em Bowen-York extrinsic
  curvature}\cite{Bowen:1979,Bowen-York:1980,Kulkarni-Shepley-York:1983}. 
All that remains is to solve a single elliptic
equation, the Hamiltonian constraint.  This approach has been used in
several variations\cite{Thornburg:1987,Cook-Choptuik-etal:1993,
  Brandt-Brugmann:1997}.

However, these numerical simplifications come at a cost.
The freely specifiable pieces have been restricted to a small subset of
all possible choices.  One therefore can generate only a subset of
all possible initial-data sets, one that might not contain the desired
astrophysically relevant initial-data sets.

Over the last few years there have been additional developments:
Post-Newtonian results have indicated that binary black hole metrics
are not conformally flat\cite{Rieth:1997,
Damour-Jaranowski-Schaefer:2000b}.
With certain restrictions on the slicing, it has also been shown that a
single stationary spinning black hole cannot be represented with a
conformally flat spatial metric
\cite{Monroe:1976,Garat-Price:2000}.  In \cite{Pfeiffer-Teukolsky-Cook:2000},
it was shown that conformally flat initial data sets for spinning
binary black holes contain an unphysical contamination.  Moreover,
computations in spherical symmetry\cite{Lousto-Price:1998} indicated
that initial-data sets depend strongly on the choice of the extrinsic
curvature and that the use of the Bowen-York extrinsic curvature might
be problematic.

Therefore it is necessary to move beyond conformally flat initial data
and to explore different choices for the extrinsic curvature.  Matzner
et al\cite{Matzner-Huq-Shoemaker:1999} proposed a non-flat conformal
metric based on the superposition of two Kerr-Schild metrics; a
solution based on this proposal was obtained in
\cite{Marronetti-Matzner:2000}. This work demonstrated the existence
of solutions to the 3D set of equations, but did not examine the data
sets in any detail.
Refs.~\cite{Gourgoulhon-Grandclement-Bonazzola:2001a,Grandclement-Gourgoulhon-Bonazzola:2001b} obtained solutions to a
similar set of equations during the computation of quasi-circular
orbits of binary black holes.  However, these works assumed conformal
flatness.

In this paper we present a code capable of solving the three
above-mentioned decompositions of the constraint equations for
arbitrary choices of the freely specifiable pieces.  This code is
based on spectral methods which have been used successfully for several
astrophysical problems (see
e.g. \cite{Bonazzola-Gourgoulhon-etal:1993,
Bonazzola-Gourgoulhon-Marck:1998,
Kidder-Scheel-etal:2000,
Gourgoulhon-Grandclement-etal:2001, 
Kidder-Scheel-Teukolsky:2001, 
Ansorg-Kleinwaechter-Meinel:2002,
Grandclement-Gourgoulhon-Bonazzola:2001b}).  Our code is described in
detail in a separate paper\cite{Pfeiffer-Kidder-etal:2002}.

 We compute solutions of the different decompositions for the non-flat
conformal metric proposed in Ref.~\cite{Matzner-Huq-Shoemaker:1999}.
Each decomposition has certain choices for the freely specifiable
pieces and boundary conditions that seem ``natural'' and which we use
in our solutions.  We compare the computed initial-data sets with each
other and with the ``standard'' conformally-flat solution using the
Bowen-York extrinsic curvature. Our major results confirm that
\begin{enumerate}
\item\label{res:1}the different decompositions generate different
physical initial-data sets for seemingly similar choices for the
freely specifiable pieces.
\item the choice of extrinsic curvature is critical.
\end{enumerate}
The first result is certainly not unexpected, but each of these
factors can cause relative differences of several per cent in 
gauge-invariant quantities like the ADM-energy.

We also find that the conformal TT/physical TT decompositions generate
initial-data sets with ADM-energies $2-3$\% higher than data sets of
the conformal thin sandwich decomposition.  We demonstrate that this
higher ADM-energy is related to the choice of the freely specifiable part
of the extrinsic curvature.  In addition, we find that the solutions
depend significantly on the boundary conditions used.

The paper is organized as follows. In the next section we describe the
three decompositions. Section \ref{sec:FreePieces} explains how we
choose the freely specifiable data within each decomposition.  In
section \ref{sec:Implementation} we describe and test our elliptic
solver.  Section \ref{sec:Results} presents our results, which we
discuss in section \ref{sec:Discussion}.

\section{Decompositions of Einstein's equations and the constraint equations}

\subsection{3+1 Decomposition}

In this paper we use the standard 3+1 decomposition of Einstein's
equations. We foliate the spacetime with $t=\mbox{const}$
hypersurfaces and write the four-dimensional metric as
\begin{equation}
  ^{(4)}ds^2=-N^2\,dt^2+\gamma_{ij}(dx^i+N^idt)(dx^j+N^jdt),
\end{equation}
where $\gamma_{ij}$ represents the induced 3-metric on the
hypersurfaces, and $N$ and $N^i$ represent the lapse function and the
shift vector, respectively.  We define the extrinsic curvature $K_{ij}$
on the slice by 
\begin{equation}\label{eq:K}
{\bf K}=-\frac{1}{2}\,{\perp}\:{\cal L}_n{^{(4)}{\bf g}}
\end{equation}
where ${}^{(4)}{\bf g}$ is the space-time metric, $n$ the unit
normal to the hypersurface, and $\perp$ denotes the projection operator
into the $t=\mbox{const}$ slice.
Einstein's
equations divide into constraint equations, which constrain the data
$(\gamma_{ij}, K^{ij})$ on each hypersurface, and into evolution
equations, which determine how the data $(\gamma_{ij}, K^{ij})$ evolve
from one hypersurface to the next.
The constraint equations are 
\begin{align}
  \label{eq:HamiltonianConstraintPhysical}
  R+K^2-K_{ij}K^{ij}&=16\pi G\rho\\
  \label{eq:MomentumConstraint}
  \nabla_j\left(K^{ij}-\gamma^{ij}K\right)&=8\pi G j^i.
\end{align}
Eq.~(\ref{eq:HamiltonianConstraintPhysical}) is called the {\em
  Hamiltonian constraint}, and Eq.~(\ref{eq:MomentumConstraint}) is
referred to as the {\em momentum constraint}.  $K=\gamma_{ij}K^{ij}$
is the trace of the extrinsic curvature, $\nabla$ and $R$ denote the
three dimensional covariant derivative operator and the Ricci scalar
compatible with $\gamma_{ij}$. $\rho$ and $j^i$ are the energy and
momentum density, respectively. Both vanish for the vacuum spacetimes
considered here.

The evolution equation for $\gamma_{ij}$ is
\begin{equation}
  \label{eq:Evolution-gamma}
  \partial_t\gamma_{ij}=-2N K_{ij}+\nabla_iN_j+\nabla_jN_i,
\end{equation}
which follows from Eq.~(\ref{eq:K}).  There is a similar albeit
longer equation for $\partial_tK_{ij}$ which we will not need in this
paper. The choices of $N$ and $N^i$ are arbitrary. One can in
principle use any lapse and shift in the evolution off the
initial-data surface, although some choices of lapse and shift are better
suited to numerical implementation than others.

Later in this paper we will often refer to the trace-free piece of 
Eq.~(\ref{eq:Evolution-gamma}).  Denote the tracefree piece of a tensor
by $\TF(.)$, and define $\gamma\equiv\det \gamma_{ij}$.  From Eq.~(\ref{eq:Evolution-gamma}) and the fact that
$\delta\ln\gamma=\gamma^{kl}\delta\gamma_{kl}$, it follows that
\begin{equation}\label{eq:Evolution-gamma-TF}
\TF(\partial_t\gamma_{ij})
=\gamma^{1/3}\partial_t\left(\gamma^{-1/3}\gamma_{ij}\right)
=-2N A_{ij}+(\Long N)_{ij}.
\end{equation}
Here $A_{ij}=K_{ij}-\frac{1}{3}\gamma_{ij}K$ denotes the trace-free
extrinsic curvature, and
\begin{equation}
  \label{eq:DefinitionLong}
  (\Long N)^{ij}\equiv 
  \nabla^iN^j+\nabla^jN^i-\frac{2}{3}\gamma^{ij}\nabla_kN^k.
\end{equation}
$\Long$ always acts on a vector, so the 'N' in $(\Long N)^{ij}$ denotes
the shift vector $N^i$ and not the lapse $N$. 

\subsection{Decomposition of the constraint equations}
\label{sec:Decomposition}

Equations~(\ref{eq:HamiltonianConstraintPhysical}) and
(\ref{eq:MomentumConstraint}) constrain four degrees of freedom of the
12 quantities $(\gamma_{ij}, K^{ij})$. However, it is not immediately
clear which pieces of $\gamma_{ij}$ and $K^{ij}$ are constrained and
which pieces can be chosen at will.  Several decompositions have been
developed to divide the 12 degrees of freedom into freely specifiable
and constrained pieces.  We will now review some properties of
the three decompositions we consider in this paper.

All three decompositions follow the York-Lichnerowicz approach and
use a conformal transformation on the
physical 3-metric $\gamma_{ij}$,
\begin{equation}
  \label{eq:ConformalMetric}
  \gamma_{ij}=\psi^4\tilde\gamma_{ij}.
\end{equation}
$\psi$ is called the {\em conformal factor}, $\tilde\gamma_{ij}$ the
{\em background metric} or {\em conformal metric}.  We will denote all
conformal quantities with a tilde. In particular, $\tilde\nabla$ is
the covariant derivative operator associated with $\tilde\gamma_{ij}$,
and $\tilde R_{ij}$ and $\tilde R$ are the Ricci tensor and Ricci
scalar of $\tilde\gamma_{ij}$.

The extrinsic curvature is split into its trace and trace-free part,
\begin{equation}
  \label{eq:SplitK_Aij}
  K^{ij}=A^{ij}+\frac{1}{3}\gamma^{ij}K.
\end{equation}
The three decompositions of the constraint equations we discuss in
this paper differ in how $A^{ij}$ is decomposed.  For each
decomposition, we discuss next the relevant equations, and describe
how we choose the quantities one has to specify before solving the
equations.  We use the conventions of \cite{Cook:2000}.

\subsubsection{Conformal TT Decomposition}

In this decomposition one first conformally transforms the traceless
extrinsic curvature,
\begin{equation}
  \label{eq:ConfTT-AijConfAij}
  A^{ij}=\psi^{-10}\tilde A^{ij},
\end{equation}
and then applies a TT decomposition with respect to the background
metric $\tilde\gamma_{ij}$:
\begin{equation}
\label{eq:ConfTT-Aij}
\tilde A^{ij}=\tilde A_{TT}^{ij}+(\tilde\Long X)^{ij}.
\end{equation}
The operator $\tilde\Long$ is defined by Eq.~(\ref{eq:DefinitionLong}) but
using the conformal metric $\tilde\gamma_{ij}$ and derivatives
associated with $\tilde\gamma_{ij}$. $\tilde A^{ij}_{TT}$ is
transverse with respect to the conformal metric, $\tilde\nabla_j\tilde
A^{ij}_{TT}=0$, and is traceless. 

Substituting Eqs.~(\ref{eq:ConfTT-AijConfAij}) and (\ref{eq:ConfTT-Aij})
into the momentum constraint (\ref{eq:MomentumConstraint}), one finds
that it reduces to an elliptic equation for $X^i$, whereas $\tilde
A^{ij}_{TT}$ is unconstrained. 

In order to specify the transverse-traceless tensor $\tilde
A^{ij}_{TT}$ one usually has to {\em construct} it from a general
symmetric trace-free tensor $\tilde M^{ij}$ by subtracting the
longitudinal piece.  As described in \cite{Cook:2000} one can
incorporate the construction of $\tilde A^{ij}_{TT}$ from $\tilde
M^{ij}$ into the momentum constraint, arriving at the following
equations:

\begin{gather}
\label{eq:ConfTT-1}
    \tilde\nabla^2\psi-\frac{1}{8}\psi\tilde R-\frac{1}{12}\psi^5K^2
    +\frac{1}{8}\psi^{-7}\tilde A_{ij}\tilde A^{ij}=-2\pi G\psi^5\rho,\\
\label{eq:ConfTT-2}
    \tildeLapLong V^i-\frac{2}{3}\psi^6\tilde\nabla^iK
    +\tilde\nabla_j\tilde M^{ij}=8\pi G\psi^{10}j^i,
\end{gather}
where $\tilde A^{ij}$ and the operator $\tildeLapLong$ are defined by
\begin{equation}
\label{eq:ConfTT-Aij_tilde}
  \tilde A^{ij}=(\tilde\Long V)^{ij}+\tilde M^{ij}
\end{equation}
and
\begin{equation}
  \label{eq:DefinitionLapLong}
  \tildeLapLong V^i\equiv\tilde\nabla_j(\tilde\Long V)^{ij}.
\end{equation}

After solving these equations for $\psi$ and $V^i$, one obtains the
physical metric $\gamma_{ij}$ from (\ref{eq:ConformalMetric}) and the
extrinsic curvature from
\begin{equation}
\label{eq:ConfTT-3}
    K^{ij}=\psi^{-10}\tilde A^{ij}+\frac{1}{3}\psi^{-4}\tilde\gamma^{ij}K. 
\end{equation}
We will refer to Eqs.~(\ref{eq:ConfTT-1}) and (\ref{eq:ConfTT-2})
together with (\ref{eq:ConfTT-Aij_tilde}), (\ref{eq:ConfTT-3}) and
(\ref{eq:ConformalMetric}) as the {\em conformal TT equations}.  In
these equations we are free to specify the background metric
$\tilde\gamma_{ij}$, the trace of the extrinsic curvature $K$, and a
symmetric traceless tensor $\tilde M^{ij}$.  The solution $V^i$ will
contain a contribution that removes the longitudinal piece from
$\tilde M^{ij}$ and the piece that solves the momentum constraint if
$\tilde M^{ij}$ were transverse-traceless.

This decomposition has been the most important in the past, since if
one chooses a constant $K$ and if one considers vacuum spacetimes then
the momentum constraint (\ref{eq:ConfTT-2}) decouples from the
Hamiltonian constraint (\ref{eq:ConfTT-1}). Moreover, if one assumes
conformal flatness and $\tilde M^{ij}=0$, it is possible to write down
analytic solutions to Eq.~(\ref{eq:ConfTT-2}), the so-called
Bowen-York extrinsic curvature.  In that case one has to deal with
only one elliptic equation for $\psi$.  The Bowen-York extrinsic
curvature can represent multiple black holes with arbitrary momenta
and spins.  One can fix boundary conditions for $\psi$ by requiring
that the initial-data slice be inversion symmetric at both
throats\cite{Misner:1963,Cook:1991}.  In that case one has to modify the
extrinsic curvature using a method of images.  We will
include initial-data sets obtained with this approach below, where we
will refer to them as {\em inversion symmetric} initial data.

Reasonable choices for the freely specifiable pieces $\tilde\gamma_{ij}$, $K$,
$\tilde M^{ij}$ will lead to an initial-data set $(\gamma_{ij},
K^{ij})$ that satisfies the constraint equations. How should we choose
all these functions in order to obtain a desired physical
configuration, say a binary black hole with given linear momenta and
spins for the individual holes?  We can gain insight into this question
by considering how the conformal TT decompositions can recover a known
solution.

Suppose we have a known solution $(\gamma_{0\,ij}, K^{ij}_0)$
of the constraint equations. Denote the trace and trace-free parts of
this extrinsic curvature by $K_0$ and $A^{ij}_0$, respectively.  If we
set 
\begin{equation}
\tilde\gamma_{ij}=\gamma_{0\,ij},\quad K=K_0,
\quad\tilde M^{ij}=A^{ij}_0
\end{equation}
then 
\begin{equation}
\psi=1,\quad V^i=0
\end{equation}
trivially solve Eqs.~(\ref{eq:ConfTT-1}-\ref{eq:ConfTT-2}). Note
that we have to set $\tilde M^{ij}$ equal to the trace-free part of the
extrinsic curvature.

Now suppose we have a guess for a metric and an extrinsic curvature,
which ---most likely--- will not satisfy the constraint equations
(\ref{eq:HamiltonianConstraintPhysical}) and
(\ref{eq:MomentumConstraint}). Set $\tilde\gamma_{ij}$ to the guess
for the metric, and set $K$ and $\tilde M^{ij}$ to the trace and
trace-free piece of the guess of the extrinsic curvature.  By solving
the conformal TT equations we can compute $(\gamma_{ij}, K^{ij})$ that
satisfy the constraint equations.  If our initial guess is ``close''
to a true solution, we will have $\psi\approx 1$ and $V^i\approx 0$,
so that $\gamma_{ij}$ and $K^{ij}$ will be close to our initial guess.

Thus one can guess a metric and extrinsic curvature as well as
possible and then solve the conformal TT equations to obtain corrected
quantities that satisfy the constraint equations.

An artifact of the conformal TT decomposition is that one has no
direct handle on the transverse traceless piece with respect to the
{\em physical} metric.  For any vector $X^i$,
\begin{equation}\label{eq:ConformalLong}
  (\Long X)^{ij}=\psi^{-4}(\tilde\Long X)^{ij}.
\end{equation}
Thus, Eqs.~(\ref{eq:ConfTT-AijConfAij}) and (\ref{eq:ConfTT-Aij}) imply
\begin{equation}\label{eq:ConfTT-Decomposition-psi}
  A^{ij}=\psi^{-10}\tilde A^{ij}_{TT}+\psi^{-6}(\Long X)^{ij}.
\end{equation}
For any symmetric traceless tensor $S^{ij}$
\begin{equation}
  \label{eq:DecompositionSymmetricTrace-Free}
  \nabla_jS^{ij}=\psi^{-10}\tilde\nabla_j\left(\psi^{10}S^{ij}\right).
\end{equation}
Therefore the first term on the right hand side of
Eq.~(\ref{eq:ConfTT-Decomposition-psi}) is transverse-traceless with
respect to the physical metric, 
\begin{equation}
  \nabla_j\left(\psi^{-10}\tilde A^{ij}_{TT}\right)=0.
\end{equation}
However, the second term on the right hand side of
Eq.~(\ref{eq:ConfTT-Decomposition-psi}) is conformally weighted.
Therefore, Eq.~(\ref{eq:ConfTT-Decomposition-psi}) does not represent the
usual TT decomposition.

\subsubsection{Physical TT Decomposition}

In this case one decomposes the physical traceless extrinsic
curvature directly:
\begin{equation}
  \label{eq:PhysTT-Aij}
  A^{ij}=A^{ij}_{TT}+(\Long X)^{ij}.
\end{equation}
As above in the conformal TT decomposition, the momentum constraint
becomes an elliptic equation for $X^i$. We can again incorporate the
construction of the symmetric transverse traceless tensor
$A^{ij}_{TT}$ from a general symmetric tensor $\tilde M^{ij}$ into the
momentum constraint. Then one obtains the {\em physical TT equations}:

\begin{gather}
\label{eq:PhysTT-1}
    \tilde\nabla^2\psi-\frac{1}{8}\psi\tilde R-\frac{1}{12}\psi^5K^2
    +\frac{1}{8}\psi^{5}\tilde A_{ij}\tilde A^{ij}=-2\pi G\psi^5\rho,\\
\label{eq:PhysTT-2}
    \tildeLapLong V^i+6(\tilde\Long V)^{ij}\tilde\nabla_j\ln\psi
    -\frac{2}{3}\tilde\nabla^iK
    +\psi^{-6}\tilde\nabla_j\tilde M^{ij}=8\pi G\psi^4j^i,
\end{gather}
where $\tilde A^{ij}$ is defined by
\begin{equation}
    \tilde A^{ij}=(\tilde\Long V)^{ij}+\psi^{-6}\tilde M^{ij}.
\end{equation}
When we have solved (\ref{eq:PhysTT-1}) and (\ref{eq:PhysTT-2}) for
$\psi$ and $V^i$, the physical metric is given by
(\ref{eq:ConformalMetric}), and the extrinsic curvature is
\begin{equation}
\label{eq:PhysTT-3}
K^{ij}=\psi^{-4}\left(\tilde A^{ij}+\frac{1}{3}\tilde\gamma^{ij}K\right).
\end{equation}

We are free to specify the background metric $\tilde\gamma_{ij}$, the
trace of the extrinsic curvature $K$, and a symmetric traceless tensor
$\tilde M^{ij}$.  As with the conformal TT equations, the solution
$V^i$ will contain a contribution that removes the longitudinal piece
from $\tilde M^{ij}$ and a piece that solves the momentum constraint
if $\tilde M^{ij}$ were transverse-traceless.

These equations can be used in the same way as the conformal TT
equations.  Guess a metric and extrinsic curvature, set
$\tilde\gamma_{ij}$ to the guess for the metric, and $K$ and $\tilde
M^{ij}$ to the trace and trace-free pieces of the guess for the
extrinsic curvature. Then solve the physical TT equations to obtain a
corrected metric $\gamma_{ij}$ and a corrected extrinsic curvature
$K^{ij}$ that satisfy the constraint equations.

The transverse traceless piece of $K^{ij}$ (with respect to
$\gamma_{ij}$) will be the transverse traceless piece of
$\psi^{-10}\tilde M^{ij}$.  One can also easily rewrite the physical
TT equations such that $\psi^{-10}\tilde M^{ij}$ can be freely chosen
instead of $\tilde M^{ij}$.  So, in this decomposition we can directly
control the TT piece of the physical extrinsic curvature.  We have
chosen to follow \cite{Cook:2000} since it seems somewhat more natural
to specify two conformal quantities, $\tilde\gamma_{ij}$ and $\tilde
M^{ij}$ than to specify one conformal and one physical quantity.

\subsubsection{Conformal thin sandwich decomposition}

The conformal and physical TT decompositions rely on a tensor
splitting to decompose the trace-free part of the extrinsic curvature.
In contrast, the conformal thin sandwich decomposition simply defines
 $A^{ij}$ by Eq.~(\ref{eq:ConfTT-AijConfAij}) and the decomposition
\begin{equation}
\label{eq:SandwichTT-3}
  \tilde A^{ij}\equiv\frac{1}{2\tilde\alpha}
\left((\tilde\Long \beta)^{ij}-\tilde u^{ij}\right),
\end{equation}
where $\tilde u^{ij}$ is symmetric and tracefree.
Eq.~(\ref{eq:SandwichTT-3}) is motivated by
Eq.~(\ref{eq:Evolution-gamma-TF}): If one evolves an initial-data set
with $A^{ij}$ of the form (\ref{eq:SandwichTT-3}) using as lapse and
shift
\begin{equation}\label{eq:SandwichTT-LapseShift}
\begin{aligned}
  N&=\psi^6\tilde\alpha,\\
  N^i&=\beta^i,
\end{aligned}
\end{equation}
then 
\begin{equation}\label{eq:TF-sandwich}
\TF(\partial_t\gamma_{ij})=\psi^4\tilde u_{ij}.
\end{equation}
Therefore, the decomposition (\ref{eq:SandwichTT-3}) is closely
related to the kinematical quantities in an evolution. Although
$\tilde\alpha$ and $\beta^i$ are introduced in the context of initial
data, one usually refers to them as the ``conformal lapse'' and ``shift''.
While the form of Eq.~(\ref{eq:SandwichTT-3}) is similar in form to
the conformal and physical TT decompositions, there are differences.
In particular, $\tilde{u}^{ij}$ is {\em not} divergenceless.

Within the {\em conformal thin sandwich decomposition}, the constraint
equations take the form:
\begin{gather}
\label{eq:SandwichTT-1}
    \tilde\nabla^2\psi-\frac{1}{8}\psi\tilde R-\frac{1}{12}\psi^5K^2
    +\frac{1}{8}\psi^{-7}\tilde A_{ij}\tilde A^{ij}=-2\pi G\psi^5\rho\\
\tildeLapLong\beta^i-(\tilde\Long\beta)^{ij}\tilde\nabla_j\ln\tilde\alpha
-\frac{4}{3}\tilde\alpha\psi^6\tilde\nabla^iK
\qquad\qquad\qquad\qquad\nonumber\\
\label{eq:SandwichTT-2}
\qquad\qquad\qquad\qquad -\tilde\alpha\tilde\nabla_j\Big(\frac{1}{\tilde\alpha}\tilde u^{ij}\Big)
=16\pi G\tilde\alpha\psi^{10}j^i
\end{gather}
Having solved Eqs.~(\ref{eq:SandwichTT-1}) and (\ref{eq:SandwichTT-2}) 
for $\psi$ and the vector $\beta^i$, one obtains the physical metric
from (\ref{eq:ConformalMetric}) and the extrinsic curvature from
\begin{equation}
    K^{ij}=\psi^{-10}\tilde A^{ij}+\frac{1}{3}\psi^{-4}\tilde\gamma^{ij}K.
\end{equation}
In this decomposition we are free to specify a conformal metric
$\tilde\gamma_{ij}$, the trace of the extrinsic curvature $K$, a
symmetric trace-free tensor $\tilde u^{ij}$ and a function $\tilde\alpha$.

It seems that the conformal thin sandwich decomposition contains
additional degrees of freedom in the form of the function
$\tilde\alpha$ and three additional unconstrained components of
$\tilde{u}^{ij}$. This is not the case.  The longitudinal piece of
$\tilde u^{ij}$ corresponds to the gauge choice of the actual shift
vector used in an evolution.  Thus $\tilde{u}^{ij}$ really only
contributes two degrees of freedom, just like $\tilde{M}^{ij}$ in the
conformal and physical TT decompositions.  Furthermore, we can reach
any {\em reasonable} physical solution $(\gamma_{ij}, K^{ij})$ with
any {\em reasonable} choice of $\tilde\alpha$; each choice of
$\tilde\alpha$ simply defines a new decomposition.  A forthcoming
article by York\cite{York:2002} will elaborate on these ideas.  Note
that for $\tilde \alpha=1/2$ we recover the conformal TT
decomposition.

Let us now turn to the question of how one should pick the freely
specifiable data in the conformal thin sandwich approach. 
We motivate our prescription again by considering how to recover a
known spacetime: Assume we are given a full four-dimensional spacetime
with 3+1 quantities $\gamma_{0\,ij}$, $K_0^{ij}$, $N_0^i$ and $N_0$.
Further assume the spacetime is stationary and the slicing is such that
$\partial_t\gamma_{ij}=\partial_tK_{ij}=0$.  An example for such a
situation is a Kerr black hole in Kerr-Schild or
Boyer-Lindquist coordinates.

Using $\partial_t\gamma_{0\,ij}=0$ in
Eq.~(\ref{eq:Evolution-gamma-TF}) yields a relation for the trace-free
extrinsic curvature
\begin{equation}
  A_0^{ij}=\frac{1}{2N_0}(\Long N_0)^{ij}.
\end{equation}
This is a decomposition of the form (\ref{eq:SandwichTT-3}) with
$\tilde u^{ij}=0$. Therefore, if we choose the freely specifiable data
for the conformal thin sandwich equations as 
\begin{equation}
\begin{aligned}
\tilde\gamma_{ij}&=\gamma_{0\,ij},&
\tilde\alpha&=N_0,\\
K&=K_0,&\tilde u^{ij}&=0,
\end{aligned}
\end{equation}
and if we use appropriate boundary conditions, then the solution of
the conformal thin sandwich equations will be $\psi=1$ and $\beta^i=N_0^i$.
As part of the solution, we obtain the shift vector needed for an
evolution to produce $\TF(\partial_t\gamma_{ij})=0$.
Not needing a guess for the trace-free extrinsic curvature, and having
the solution $\beta^i$ automatically provide an initial shift for
evolution, make the conformal thin sandwich equations very attractive.

In order to generate initial-data slices that permit an evolution with
zero time derivative of the conformal metric --- a highly desirable
feature for quasi-equilibrium data, or for a situation with holes
momentarily at rest --- one can proceed as follows: Set
$\tilde\gamma_{ij}$ and $K$ to the guess for the metric and trace of
extrinsic curvature, respectively. Set $\tilde\alpha$ to the
lapse function that one is going to use in the evolution, and set
$\tilde u^{ij}=0$. If these guesses are good, the conformal factor
$\psi$ will be close to 1, and $N=\psi^6\tilde\alpha$ as well as
$N^i=\beta^i$ give us the actual lapse function and shift vector to use in
the evolution.

\section{Choices for the freely specifiable data}
\label{sec:FreePieces}

\subsection{Kerr-Schild coordinates}
\label{sec:Choices:Kerr-Schild}

We base our choice for the freely specifiable data on a superposition of two
Kerr black holes in Kerr-Schild coordinates. In this section we
describe the Kerr-Schild solution and collect necessary
equations. We also describe how we compute the 3-metric, lapse, shift
and extrinsic curvature for a boosted black hole with arbitrary spin.

A Kerr-Schild metric is given by
\begin{equation}\label{eq:KerrSchild}
  g_{\mu\nu}=\eta_{\mu\nu}+2Hl_\mu l_\nu,
\end{equation}
where $\eta_{\mu\nu}$ is the Minkowski metric, and $l_\mu$ is a
null-vector with respect to both the full metric and the Minkowski
metric: $g^{\mu\nu}l_\mu l_\nu=\eta^{\mu\nu}l_\mu l_\nu=0$.  The
3-metric, lapse and shift are
\begin{align}
  \label{eq:KerrSchild-gamma}
  \gamma_{ij}&=\delta_{ij}+2Hl_il_j,\\
  N&=(1+2Hl^tl^t)^{-1/2},\\
\label{eq:KerrSchild-beta}
  N^i&=-\frac{2Hl^tl^i}{1+2Hl^tl^t}.
\end{align}

For a black hole at rest at the origin with mass $M$ and angular
momentum $M\vec a$, one has
\begin{align}
  H&=\frac{Mr^3}{r^4+(\vec a\cdot\vec x)^2},\\
  l^{\mbox{\footnotesize rest}}_\mu&=(1, \vec l_{\mbox{\footnotesize rest}}),\\
  \vec l_{\mbox{\footnotesize rest}}\;\,&=\frac{r\vec x-\vec a\times\vec x+(\vec a\cdot\vec x)\vec a/r}
              {r^2+a^2},
\end{align}
with
\begin{equation}
  r^2=\frac{\vec x^2-\vec a^2}{2}
       +\left(\frac{(\vec x^2- \vec a^2)^2}{4}
       +(\vec a\cdot\vec x)^2\right)^{1/2}.
\end{equation}
For a nonrotating black hole with $\vec a=0$, $H$ has a pole at the
origin, whereas for rotating black holes, $r$ has a ring singularity.
We will therefore have to excise from the computational domain a
region close to the center of the Kerr-Schild black hole.

Under a boost, a Kerr-Schild coordinate system transforms into a
Kerr-Schild coordinate system. Applying a Lorentz transformation with
boost velocity $v^i$ to $l^{\mbox{\footnotesize rest}}_\mu$, we obtain
the null-vector $l_\mu$ of the boosted Kerr-Schild coordinate system.
Eqs.~(\ref{eq:KerrSchild-gamma}-\ref{eq:KerrSchild-beta}) give then
the boosted 3-metric, lapse, and shift.  Since all time-dependence is
in the uniform motion, evolution with lapse $N$ and shift $N^i$ yields
$\partial_t\gamma_{ij}=-v^k\partial_k\gamma_{ij}$, and from
Eq.~(\ref{eq:Evolution-gamma}) one can compute the extrinsic curvature
\begin{equation}\label{eq:KerrSchild-Kij}
  K_{ij}=\frac{1}{2N}\left(v^k\partial_k\gamma_{ij}
         +\nabla_{\!i} N_j+\nabla_{\!j} N_i\right).
\end{equation}

If this initial-data set is evolved with the shift $N^i$, the black
hole will move through the coordinate space with velocity $v^i$.
However, if the evolution uses the shift vector $N^i+v^i$, the
coordinates will move with the black hole, and the hole will be at
rest in coordinate space.  The spacetime is nonetheless different from
a Kerr black hole at rest. The ADM-momentum will be $P^i_{ADM}=\gamma
M v^i$, where $M$ is the rest-mass of the hole and $\gamma=(1-\vec
v^2)^{-1/2}$.

\subsection{Freely specifiable pieces}

We want to generate initial data for a spacetime containing two black
holes with masses $M_{\!A,B}$, velocities $\vec v_{\!A,B}$ and spins
$M_{\!A}\vec a_A$ and $M_{\!B}\vec a_B$.

We follow the proposal of Matzner et al
\cite{Matzner-Huq-Shoemaker:1999, Marronetti-Matzner:2000} and base
our choices for the freely specifiable choices on two Kerr-Schild
coordinate systems describing two individual black holes.  The first
black hole with label A has an associated Kerr-Schild coordinate
system with metric
\begin{equation}
  \label{eq:KerrSchild-HoleA-gamma}
\gamma_{A\,ij}=\delta_{ij}+2H_{\!A}\,l_{A\,i}\,l_{A\,j},
\end{equation}
and with an extrinsic curvature $K_{\!A\,ij}$, a lapse $N_{\!A}$ and a
shift $N^i_A$. The trace of the extrinsic curvature is $K_A$. All
these quantities can be computed as described in the previous section,
\ref{sec:Choices:Kerr-Schild}. The second black hole has a similar set
of associated quantities which are labeled with the letter B.

For all three decompositions, we need to choose a conformal metric and
the trace of the extrinsic curvature. We choose
\begin{gather}
  \label{eq:BinaryKerrSchild-gamma}
  \tilde\gamma_{ij}=\delta_{ij}+2H_{\!A}\,l_{A\,i}\,l_{A\,j}
  +2H_{\!B}\,l_{B\,i}\,l_{B\,j}\\
K=K_{\!A}+K_B\label{eq:BinaryKerrSchild-K}
\end{gather}
The metric is singular at the center of each hole. Therefore we have
to excise spheres around the center of each hole from the
computational domain.  We now specify for each decomposition the
remaining freely specifiable pieces and boundary conditions.

\subsubsection{Conformal TT and physical TT decompositions}

For the conformal TT and physical TT decompositions we will be solving
for a correction to our guesses. 
As guess for the trace-free extrinsic curvature, we use a superposition
\begin{equation}
  \label{eq:BinaryKerrSchild-Mij}
  \tilde M^{ij}=\left(K^{(i}_{A\,k}+K^{(i}_{B\,k}
    -\frac{1}{3}\delta^{(i}_k(K_A+K_B)\right)\tilde\gamma^{j)k}.
\end{equation}
$\tilde M^{ij}$ is symmetric and trace-free with respect to the
conformal metric, $\tilde\gamma_{ij}\tilde M^{ij}=0$.  Solving for a
correction only, we expect that $\psi\approx 1$ and $V^i\approx 0$,
hence we use Dirichlet boundary conditions
\begin{equation}
  \label{eq:BC-ConfTT-PhysTT}
  \psi=1, \qquad V^i=0.
\end{equation}

\subsubsection{Conformal thin sandwich}

For conformal thin sandwich, we restrict the discussion to either two
black holes at rest, or in a quasi-circular orbit in corotating
coordinates.  In these cases, one expects small or even vanishing
time-derivatives, $\partial_t\gamma_{ij}\approx 0$, and so
Eq.~(\ref{eq:TF-sandwich}) yields the simple choice
\begin{equation}\label{eq:CTS-uij}
\tilde u^{ij}=0.
\end{equation}

The conformal 3-metric and the trace of the extrinsic curvature are
still given by Eqs.~(\ref{eq:BinaryKerrSchild-gamma}) and
(\ref{eq:BinaryKerrSchild-K}). Orbiting black holes in a corotating
frame will not move in coordinate space, therefore we do not boost the
individual Kerr-Schild metrics in this decomposition: $v^i_{A/B}=0$.
The lapse functions $N_{A/B}$ and the shifts $N^i_{A/B}$ are also for
unboosted Kerr-Schild black holes.
   
We use Dirichlet boundary conditions:
\begin{subequations}\label{eq:BC-sandwich}
\begin{align}
\psi&=1                            &&\mbox{all boundaries}\\
\label{eq:Sandwich-BC2}
\beta^i&=N_{\!A}^i                 &&\mbox{sphere inside hole A}\\
\beta^i&=N_B^i                     &&\mbox{sphere inside hole B}\\
\label{eq:Sandwich-BC4}
\beta^i&=\vec\Omega\times\vec r    &&\mbox{outer boundary}
\end{align}
\end{subequations}

Eq.~(\ref{eq:Sandwich-BC4}) ensures that we are in a corotating
reference frame; the cross-product is performed in flat space, and
$\vec\Omega=0$ corresponds to two black holes at rest.  Close to the
holes we force the shift to be the shift of a single black hole in the
hope that this choice will produce a hole that is at rest in
coordinate space.

For the conformal lapse we use
\begin{equation}\label{eq:Sandwich-N1}
  \tilde\alpha=N_{\!A}+N_B-1
\end{equation}
or
\begin{equation}\label{eq:Sandwich-N2}
  \tilde \alpha=N_{\!A}\;N_B.
\end{equation}
The first choice of $\tilde\alpha$ follows the philosophy of adding
quantities of each individual hole. However, $\tilde\alpha$ of
Eq.~(\ref{eq:Sandwich-N1}) becomes negative sufficiently close to the
center of each hole and is therefore a bad choice if the excised
spheres are small. The choice (\ref{eq:Sandwich-N2}) does not change
sign and has at large distances the same behavior (same $1/r$ term) as
(\ref{eq:Sandwich-N1}).

\section{Numerical Implementation}
\label{sec:Implementation}

We implemented an elliptic solver that can solve all three
decompositions we described above in complete generality.  The solver
uses domain decomposition and can handle nontrivial topologies. It is
based on pseudospectral collocation, that is, it expresses the
solution in each subdomain as an expansion in basis functions.  This
elliptic solver is described in detail in a separate paper
\cite{Pfeiffer-Kidder-etal:2002}.

\begin{figure}
  \includegraphics[scale=0.33, angle=90]{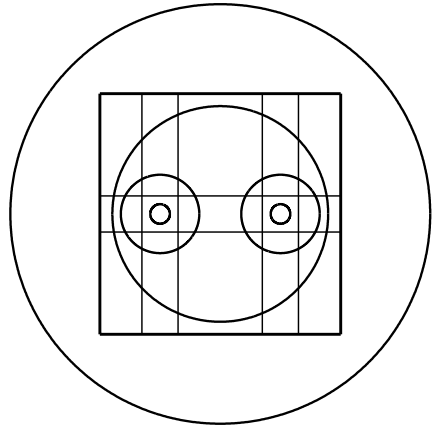}
    \caption{\label{fig:Domains}
      Structure of domains. Spherical shells around each excised
      sphere are surrounded by 43 rectangular blocks and another
      spherical shell.  The rectangular blocks touch each other and
      overlap with all three spherical shells.}
\end{figure}

From the computational domain we excise two spheres containing the
singularities of the Kerr-Schild metric close to the center of each
hole.  Around each of the excised spheres we place a spherical shell.
These shells are patched together with $5\times 3\times 3=45$
rectangular blocks, with the two blocks at the location of the spheres
removed. Around these 43 blocks, another spherical shell is placed
that extends far out, typically to an outer radius of $10^7M$.  In the
rectangular blocks, we expand in Chebyshev polynomials, while in the
spheres we use Chebyshev polynomials radially and spherical harmonics
for the angular variables.  This setup is depicted in Fig.~\ref{fig:Domains}.

The domain decomposition in Fig.~\ref{fig:Domains} is fairly
complicated. Even if the shells were made as large as possible, they do
not cover the full computational domain when the excised spheres are
close together.  Thus additional subdomains are needed in any case.
Choosing the 43 cubes as depicted allows for relatively small inner
shells and for a relatively large inner radius of the outer shell.
Thus each shell covers a region of the computational domain in
which the angular variations of the solution are fairly low, allowing
for comparatively few angular basis-functions.

The code can handle a general conformal metric. In principle, the user
needs to specify only $\tilde\gamma_{ij}$. Then the code computes
$\tilde\gamma^{ij}$, and ---using numerical derivatives--- the Christoffel
symbols, Ricci tensor and Riemann scalar.  For the special case of the
Kerr-Schild metric of a single black hole and the superposition of two
Kerr-Schild metrics, Eq.~(\ref{eq:BinaryKerrSchild-gamma}), we compute
first derivatives analytically and use numerical derivatives only to
compute the Riemann tensor.

The solver implements Eqs.~(\ref{eq:ConfTT-1}) and (\ref{eq:ConfTT-2})
for the conformal TT decomposition, Eqs.~(\ref{eq:PhysTT-1}) and
(\ref{eq:PhysTT-2}) for the physical TT decomposition, and
Eqs.~(\ref{eq:SandwichTT-1}) and (\ref{eq:SandwichTT-2}) for the
conformal thin sandwich decomposition.

After solving for $(\psi, V^i)$ [conformal TT and physical TT], or
$(\psi, \beta^i)$ [thin sandwich] we compute the physical metric
$\gamma_{ij}$ and the physical extrinsic curvature $K^{ij}$ of the
solution.  Utilizing these physical quantities $(\gamma_{ij},
K^{ij})$, we implement several analysis tools.  We evaluate the
constraints in the form of
Eq.~(\ref{eq:HamiltonianConstraintPhysical}) and
(\ref{eq:MomentumConstraint}), we compute ADM-quantities and we search
for apparent horizons. Note that these analysis tools are completely
independent of the particular decomposition; they rely only on
$\gamma_{ij}$ and $K^{ij}$.

Next we present tests ensuring that the various systems of equations
are solved correctly.  We also include tests of the
analysis tools showing that we can indeed compute constraints,
ADM-quantities and apparent horizons with good accuracy.

\subsection{Testing the conformal TT and physical TT decompositions}

We can test the solver by conformally distorting a known solution of
the constraint equations.  Given a solution to the constraint
equations $(\gamma_{0\,ij}, K^{ij}_0)$ pick functions
\begin{gather}
  \Psi>0,\qquad  W^i
\end{gather}
and set
\begin{align}
  \label{eq:Test-gamma}
  \tilde\gamma_{ij}&=\Psi^{-4}\gamma_{0\,ij},\\
  \label{eq:Test-K}
  K\;&=K_0,
\end{align}
and 
\begin{align}\label{eq:Test-ConfTT-M}
\tilde M^{ij}&=\Psi^{10}\left(K_0^{ij}-\frac{1}{3}\gamma_0^{ij}K_0\right)
                 -\Psi^4(\Long_0W)^{ij}
\intertext{for conformal TT or}
\label{eq:Test-PhysTT-M}
\tilde M^{ij}&=\Psi^{10}\left(K_0^{ij}-\frac{1}{3}\gamma_0^{ij}K_0
                 -(\Long_0W)^{ij}\right)
\end{align}
for physical TT.
With these freely specifiable pieces and appropriate boundary
conditions, a solution of the conformal TT equations
(\ref{eq:ConfTT-1}), (\ref{eq:ConfTT-2}) or the physical TT equations
(\ref{eq:PhysTT-1}), (\ref{eq:PhysTT-2}) will be
\begin{align}
  \psi&=\Psi\\
  V^i&=W^i.
\end{align}
From Eq.~(\ref{eq:ConformalMetric}) we recover our initial metric
$\gamma_{0\,ij}$, and from Eq.~(\ref{eq:ConfTT-3}) [conformal TT] or
Eq.~(\ref{eq:PhysTT-3}) [physical TT] we recover the extrinsic
curvature $K^{ij}_0$.

In our tests we used the particular choices
\begin{gather}
\label{eq:Psi}
  \Psi=1+\frac{8(r-2)}{36+x^2+0.9y^2+1.3(z-1)^2}\\
\label{eq:W^i}
  W^i=\frac{50(r-2)}{(6^4+r^4)} (-y, x, 1).
\end{gather}

\begin{figure}
\centerline{\includegraphics[scale=0.38]{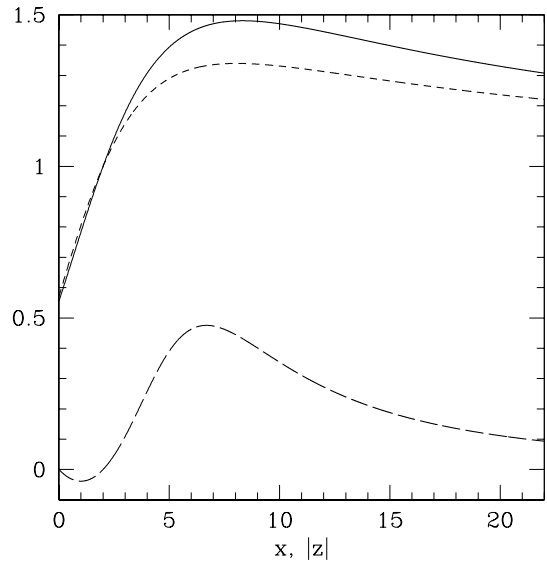}}
\caption{\label{fig:Testing-PlotDistortion}
Plot of the functions $\Psi$ and $W^i$ from Eqs.~(\ref{eq:Psi}) 
  and (\ref{eq:W^i}).  The solid line depicts $\Psi$ along the
  positive $x$-axis, the short dashed line depicts $\Psi$ along the
  negative $z$-axis. The long dashed line is a plot of $W^y$ along the
  positive $x$-axis.}
\end{figure}

These functions are plotted in Fig.~\ref{fig:Testing-PlotDistortion}.
$\Psi$ varies between 0.8 and 1.5, $W^i$ varies between $\pm 0.5$, and
both take their maximum values around distance $\sim 7$ from the
center of the hole.  We used for $(\tilde\gamma_{0\,ij}, K^{ij}_0)$ a
single, boosted, spinning black hole in Kerr-Schild coordinates.

\begin{figure}
  \centerline{\includegraphics[scale=0.38]{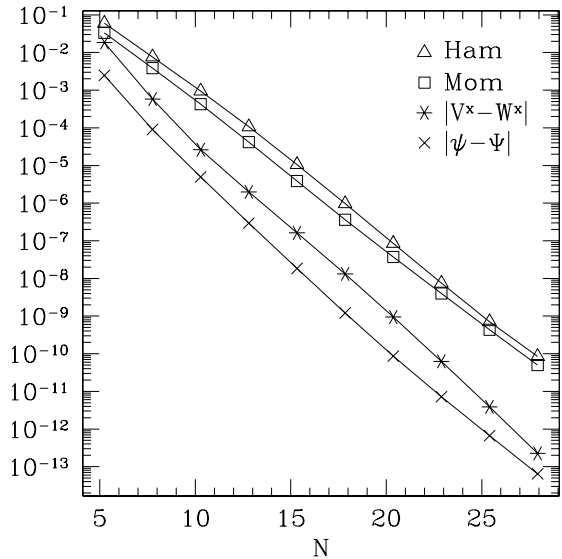}}
\caption{\label{fig:Testing-ConfTT}
  Testing the solver for the conformal TT decomposition.
  Eqs.~(\ref{eq:ConfTT-1}-\ref{eq:ConfTT-2}) with freely specifiable
  data given by Eqs.~(\ref{eq:Test-gamma}-\ref{eq:Test-ConfTT-M}) are
  solved in a single spherical shell with $1.5M<r<10M$. $N$ is the
  cube root of the total number of unknowns.  Plotted are the L2-norms
  of $\psi-\Psi$, $V^x-W^x$, and the residuals of Hamiltonian and
  momentum constraints, Eqs.~(\ref{eq:HamiltonianConstraintPhysical})
  and (\ref{eq:MomentumConstraint}).  }
\end{figure}

Figure~\ref{fig:Testing-ConfTT} shows results of testing the conformal
TT decomposition on a single spherical shell. The numerical solution
$(\psi, V^i)$ converges to the analytic solutions $(\Psi, W^i)$
exponentially with the number of basis functions as expected for a
properly constructed spectral method. Moreover, the reconstructed
metric and extrinsic curvature satisfy the constraints. 

\begin{figure}
\centerline{\includegraphics[scale=0.38]{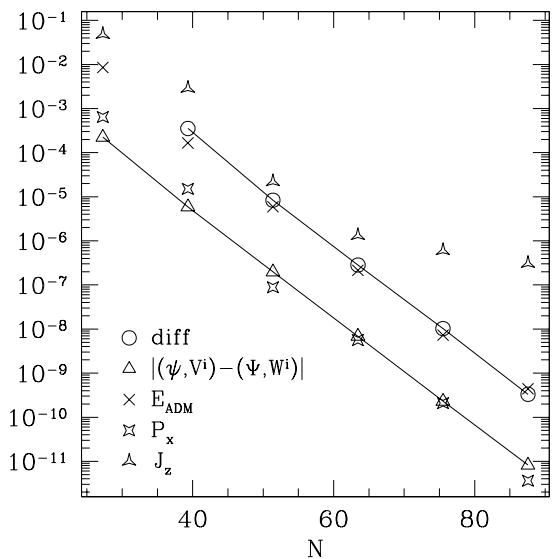}}
\caption{\label{fig:Testing-PhysTT}
  Physical TT decomposition with domain decomposition.
  Eqs.~(\ref{eq:PhysTT-1}), (\ref{eq:PhysTT-2}) with freely
  specifiable data given by Eq.~(\ref{eq:Test-gamma}, \ref{eq:Test-K},
  \ref{eq:Test-PhysTT-M}) are solved in multiple domains (one inner
  spherical shell, 26 rectangular blocks, one outer spherical shell).
  $N$ is the cube root of the total number of grid-points.  {\em diff}
  denotes the L2-norm of the difference of the solution and the
  solution at next lower resolution.  Triangles denote the L2-norm of
  the difference to the analytic solution.  The remaining symbols
  denote the errors of numerically extracted ADM-quantities. }
\end{figure}
 
Now we test the solver for the physical TT decomposition, and
demonstrate that we can correctly deal with multiple domains.  In this
example the computational domain is covered by an inner spherical
shell extending for $1.5M\le r\le 10M$.  This shell is surrounded by
26 rectangular blocks that overlap with the shell and extend out to
$x,y,z=\pm 25M$.  Finally another spherical shell covers the region
$20M\le r\le 10^6M$.  As can be seen in Fig.~\ref{fig:Testing-PhysTT},
the solution converges again exponentially.

For realistic cases we do not know the analytic solution and therefore
need a measure of the error.  Our major tool will be the change in
results between different resolution. In particular we consider the
$L_2$ norm of the point-wise differences of the solution at some resolution
and at the next lower resolution. This diagnostic is labeled by circles
in Fig.~\ref{fig:Testing-PhysTT}.  Since the solution converges
exponentially, these circles essentially give the error of the {\em
  lower} of the two resolutions.

In addition to testing the equations, this example tests domain
decomposition and the integration routines for the ADM quantities.
The ADM quantities are computed by the standard integrals at infinity
in Cartesian coordinates,
\begin{align}\label{eq:EADM}
E_{ADM}&=\frac{1}{16\pi}\int_{\infty} 
  \left(\gamma_{ij,j}-\gamma_{jj,i}\right)\,d^2S_i,\\
\label{eq:JADM}
J_{(\xi)}&=\frac{1}{8\pi}\int_{\infty}
  \left(K^{ij}-\gamma^{ij}K\right)\xi_j\,d^2S_i.
\end{align}
For the $x$-component of the linear ADM-momentum, $\xi=\hat e_x$ in
Eq.~(\ref{eq:JADM}).  The choice $\xi=x\hat e_y-y\hat e_x$ yields the
$z$-component of the ADM-like angular momentum as defined by York
\cite{York:1979}.  Since the space is asymptotically flat
there is no distinction between upper and lower indices in
Eqs.~(\ref{eq:EADM}) and (\ref{eq:JADM}). Note that
Eq.~(\ref{eq:EADM}) reduces to the familiar monopole term
\begin{equation}
-\frac{1}{2\pi}\int_{\infty}\partial_r\psi\,dA
\end{equation}
{\em only} for quasi-isotropic coordinates.
Our outer domain is large, but since
it does not extend to infinity, we extrapolate $r\to\infty$.

For a Kerr black hole with mass $M$ and spin $\vec a$, that is boosted
to velocity $\vec v$, the ADM-quantities will be
\begin{align}
  E_{ADM}&=\gamma M,\\
  \vec P_{ADM}&=\gamma M\vec v,\\
\label{eq:J-boostedBH}
  \vec J_{ADM}&=\left[\gamma\vec a-(\gamma-1)\frac{(\vec a\,\vec v)\,\vec v}{\vec  v\,^2}\right]M,
\end{align}
where $\gamma=(1-\vec v\,^2)^{-1/2}$ denotes the Lorentz factor.
Eq.~(\ref{eq:J-boostedBH}) reflects the fact that under a boost, the
component of the angular momentum perpendicular to the boost-direction
is multiplied by $\gamma$.

The example in Fig.~\ref{fig:Testing-PhysTT} uses 
$\vec v=(0.2, 0.3, 0.4)$, and $\vec a=(-1/4, 1/4, 1/6)M$.  The
evaluation of the angular momentum $J_z$ seems to be less accurate
since our current procedure to extrapolate to infinity magnifies
roundoff. We plan to improve this in a future version of the code.
Until then we seem to be limited to an accuracy of $\sim 10^{-6}$.

\subsection{Testing conformal thin sandwich equations}

The previous decompositions could be tested with a conformally
distorted known solution.  In order to test the conformal thin
sandwich decomposition we need to find an analytic decomposition of
the form (\ref{eq:SandwichTT-3}).  To do this, we
start with a stationary solution of Einstein's equations and boost it with
uniform velocity $v^i$.  Denote the metric, extrinsic curvature, lapse
and shift of this boosted spacetime by $\tilde\gamma_{0\,ij}$,
$K^{ij}_0=A^{ij}_0 +\frac{1}{3}\gamma^{ij}_0K_0$, $N_0$ and $N^i_0$,
respectively. Since we boosted the static solution, we will {\em not}
find $\partial_t\gamma_{ij}=0$ if we evolve it with the shift $N^i_0$.
However, all time-dependence of this spacetime is due to the uniform
motion, so in the comoving reference frame specified by the
shift $N^i_0+v^i$, we will find $\partial_t\gamma_{ij}=0$. In this
case, Eq.~(\ref{eq:Evolution-gamma-TF}) yields 
\begin{equation}\label{eq:Aij-boost}
A^{ij}_0=\frac{1}{2N_0}(\Long(N_0+v))^{ij}.
\end{equation}
If we choose $\tilde \alpha=N_0$ and
$\tilde u^{ij}=0$, the thin sandwich equations (\ref{eq:SandwichTT-1})
and (\ref{eq:SandwichTT-2}) will thus be solved by $\psi=1$ and
$\beta^i=N^i_0+v^i$.  Similar to the conformal TT and physical TT
decomposition above, we can also conformally distort the metric
$\gamma_{0\,ij}$.  Furthermore, we can consider nonvanishing $\tilde
u^{ij}$. We arrive at the following method to test the solver for the
conformal thin sandwich equations:

Given a boosted version of a stationary solution with shift $N_0^i$,
lapse $N_0$, 3-metric $\gamma_{0\,ij}$, trace of extrinsic curvature
$K_0$, and boost-velocity $v^i$. Pick any functions
\begin{gather}
\Psi>0\\
W^i
\end{gather}
and set
\begin{align}
\label{eq:Test-Sandwich-gamma}
\tilde\gamma_{ij}&=\Psi^{-4}\gamma_{0\,ij}\\
K\;&=K_0\\
\tilde\alpha\;\,&=\Psi^{-6}N_0\\
\label{eq:Test-Sandwich-u}
\tilde u^{ij}&=\Psi^4(\Long_0 W)^{ij}
\end{align}
Then a solution to the thin sandwich equations
(\ref{eq:SandwichTT-1}-\ref{eq:SandwichTT-2}) will be
\begin{align}
\psi&=\Psi\label{eq:psi-static-boosted}\\
\beta^i&=N_0^i+v^i+W^i\label{eq:beta-static-boosted}
\end{align}
assuming boundary conditions respecting this solution.

\begin{figure}
\centerline{\includegraphics[scale=0.38]{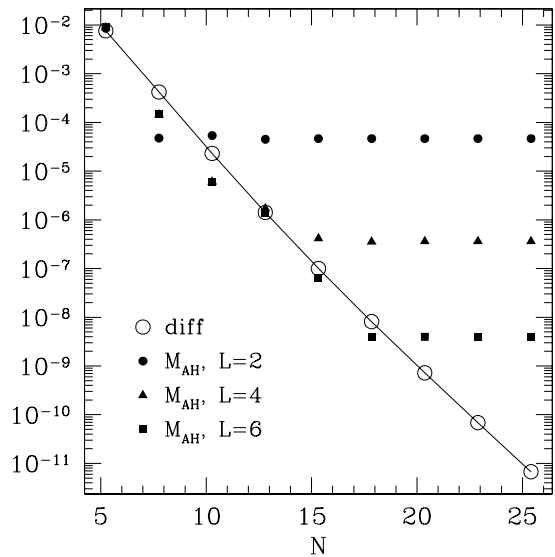}}
\caption{\label{fig:Testing-Sandwich}
  Testing thin sandwich decomposition with apparent horizon searches.
  Equations~(\ref{eq:SandwichTT-1}) and (\ref{eq:SandwichTT-2}) with
  freely specifiable data given by 
  Eqs.~(\ref{eq:Test-Sandwich-gamma})--(\ref{eq:Test-Sandwich-u}) are solved
  in a single spherical shell.  $N$ and {\em diff} as in 
  Fig.~\ref{fig:Testing-PhysTT}.  Apparent horizon searches with different
  surface expansion order $L$ were performed, and the errors of the 
  apparent horizon mass $M_{AH}$ are plotted. }
\end{figure}

Figure~\ref{fig:Testing-Sandwich} shows results of this test for a
single spherical shell and a Kerr black hole with $\vec v=(0.2,
-0.3, 0.1)$, $\vec a=(0.4, 0.3, 0.1)M$. The solution converges to the
expected analytical result exponentially.  In addition, apparent
horizon searches were performed. For the numerically found apparent
horizons, the apparent horizon area $A_{AH}$ as well as the apparent
horizon mass
\begin{equation}
  \label{eq:M_AH}
  M_{AH}=\sqrt{\frac{A_{AH}}{16\pi}}
\end{equation}
were computed. The figure compares $M_{AH}$ to the expected value
\begin{equation}
M\left(\frac{1}{2}+\frac{1}{2}\sqrt{1-\frac{\vec a^2}{M^2}}\;\right)^{1/2}.
\end{equation}
As described in \cite{Baumgarte-Cook-etal:1996,
  Pfeiffer-Teukolsky-Cook:2000}, the apparent horizon finder expands
the apparent horizon surface in spherical harmonics up to a fixed
order $L$.  For fixed $L$, the error in the apparent horizon mass is
dominated by discretization error of the elliptic solver at low
resolution $N$.  As $N$ is increased, the discretization error of the
elliptic solver falls below the error due to finite $L$.  Then the
error in $M_{AH}$ becomes independent of $N$. Since the expansion in
spherical harmonics is {\em spectral}, the achievable resolution
increases exponentially with $L$.  Note that for exponential
convergence it is necessary to position the rays in the apparent
horizon finder at the abscissas of Gauss-Legendre integration.

\subsection{Convergence of binary black hole solutions}

\begin{figure}
  \centerline{\includegraphics[scale=0.38]{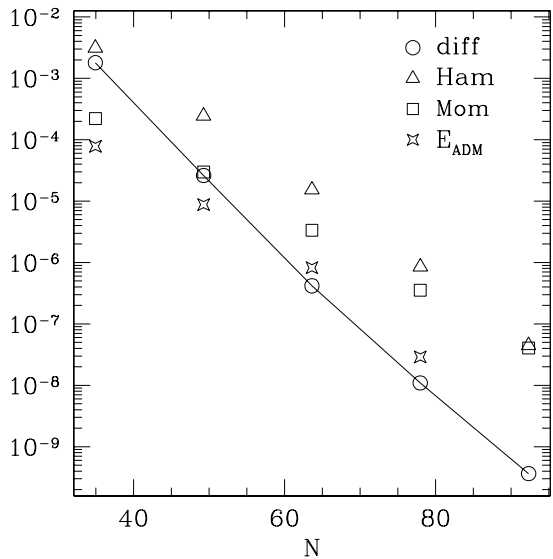}}
  \caption{\label{fig:Convergence-09F}
    Binary black hole with conformal TT decomposition. The residuals
    of several quantities are plotted as a function of the cube root
    of the total number of grid points. {\em diff} as in 
    Fig.~\ref{fig:Testing-PhysTT}, {\em Ham} and {\em Mom} are the
    residuals of Hamiltonian and momentum constraints.  $E_{ADM}$
    denotes the difference between ADM-energy at resolution $N$ and
    ADM-energy at highest resolution.}
\end{figure}

Figure~\ref{fig:Convergence-09F} present the convergence of the solver
in the binary black hole case. In this particular example, the
conformal TT equations were solved for two black holes at rest with
coordinate separation of $10M$.  The computational domain is structured
as in Fig.~\ref{fig:Domains}.  The excised spheres have radius
$r_{exc}=2M$, the inner spherical shells extend to radius $4M$. The
rectangular blocks cover space up to $x,y,z=\pm 25M$, and the outer
spherical shell extending from inner radius $20M$ to an outer radius
of $R=10^7M$.

We do not use fall-off boundary conditions at the outer boundary; we
simply set $\psi=1$ there.  This limits the computations presented in
this paper to an accuracy of order $1/R\sim 10^{-7}$. 
Figure~\ref{fig:Convergence-09F} shows that even for the next to
highest resolution ($N\approx 80$) the solution will be limited by the
outer boundary condition.  All results presented in the following
section are obtained at resolutions around $N\approx 80$.  If the need
arises to obtain solutions with higher accuracy, one can easily change
to a fall-off boundary condition, or just move the outer boundary
further out.

\section{Results}
\label{sec:Results}

The purpose of this paper is to compare the initial-data sets
generated by different decompositions using simple choices for
the freely specifiable pieces in each decomposition.  We solve 

$\bullet$ {\bf ConfTT:} Conformal TT equations (\ref{eq:ConfTT-1})
and (\ref{eq:ConfTT-2}) with freely specifiable pieces and boundary
conditions given by Eqs.~(\ref{eq:BinaryKerrSchild-gamma}),
(\ref{eq:BinaryKerrSchild-K}), (\ref{eq:BinaryKerrSchild-Mij})
and (\ref{eq:BC-ConfTT-PhysTT}).
  
$\bullet$ {\bf PhysTT:} Physical TT equations (\ref{eq:PhysTT-1})
and (\ref{eq:PhysTT-2}) with same freely specifiable pieces and
boundary conditions as ConfTT.
  
$\bullet$ {\bf CTS:} Conformal thin sandwich
equations (\ref{eq:SandwichTT-1}) and (\ref{eq:SandwichTT-2}) with
freely specifiable pieces and boundary conditions given by
Eqs.~(\ref{eq:BinaryKerrSchild-gamma}), (\ref{eq:BinaryKerrSchild-K}),
(\ref{eq:CTS-uij}) and (\ref{eq:BC-sandwich}). The lapse
$\tilde\alpha$ is given by either Eq.~(\ref{eq:Sandwich-N1}), or by
Eq.~(\ref{eq:Sandwich-N2}).
  
We will apply the terms ``ConfTT'', ``PhysTT'' and ``CTS'' only to
these particular choices of decomposition, freely specifiable pieces
and boundary conditions.  When referring to different freely
specifiable pieces, or a decomposition in general, we will not use
these shortcuts.  If we need to distinguish between the two choices of
$\tilde\alpha$ in CTS, we will use ``CTS-add'' for the additive lapse
Eq.~(\ref{eq:Sandwich-N1}) and ``CTS-mult''for the multiplicative
lapse Eq.~(\ref{eq:Sandwich-N2}).
Below in section \ref{sec:mConfTT} we will also introduce as a forth
term ``mConfTT''.

\subsection{Binary black hole at rest}
\label{sec:ResultsBHatRest}

We first examine the simplest possible configuration: Two black holes
at rest with equal mass, zero spin, and with some fixed proper
separation between the apparent horizons of the holes.  We solve
\begin{itemize}
\item ConfTT
\item PhysTT
\item CTS (with both choices of $\tilde\alpha$).  
\end{itemize}
In the comparisons, we also include inversion symmetric conformally
flat initial data obtained with the conformal-imaging formalism.

We excised spheres with radius $r_{exc}=2M$, which is close to the
event horizon for an individual Eddington-Finkelstein black hole.
This results in the boundary conditions being imposed close to, but
within the apparent horizons of the black holes.  
The centers of the excised spheres have
a coordinate separation of $d=10M$.

We now discuss the solutions.  The conformal factor $\psi$ is very
close to $1$ for each of the three decompositions. It deviates from
$1$ by less than 0.02, indicating that a conformal metric based of a
superposition of Kerr-Schild metrics does not deviate far from the
constraint surface.

\begin{figure}
\centerline{\includegraphics[scale=.38]{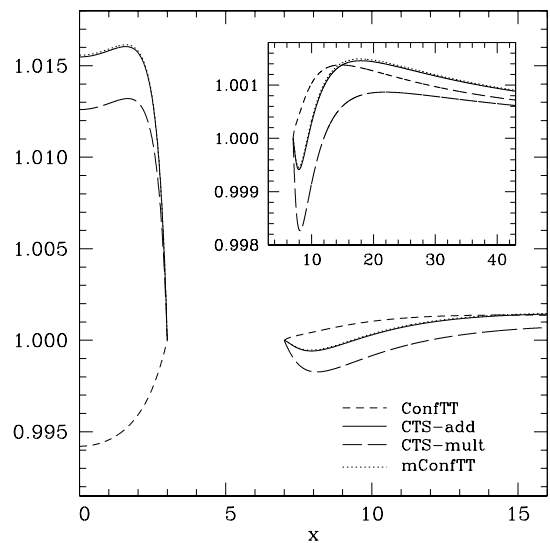}}
\caption{\label{fig:Compare-cuts}
  The conformal factor $\psi$ along the axis connecting the holes for
  several decompositions. $x$ measures the distance from the center of
  mass, so that the excised sphere is located between $3<x<7$.
  mConfTT is explained below in section \ref{sec:mConfTT}. The
  solution of PhysTT is not plotted since it is within the line
  thickness of ConfTT.  The insert shows an enlargement for large $x$.
  }
\end{figure}

\begin{figure}
\centerline{
\includegraphics[scale=0.23]{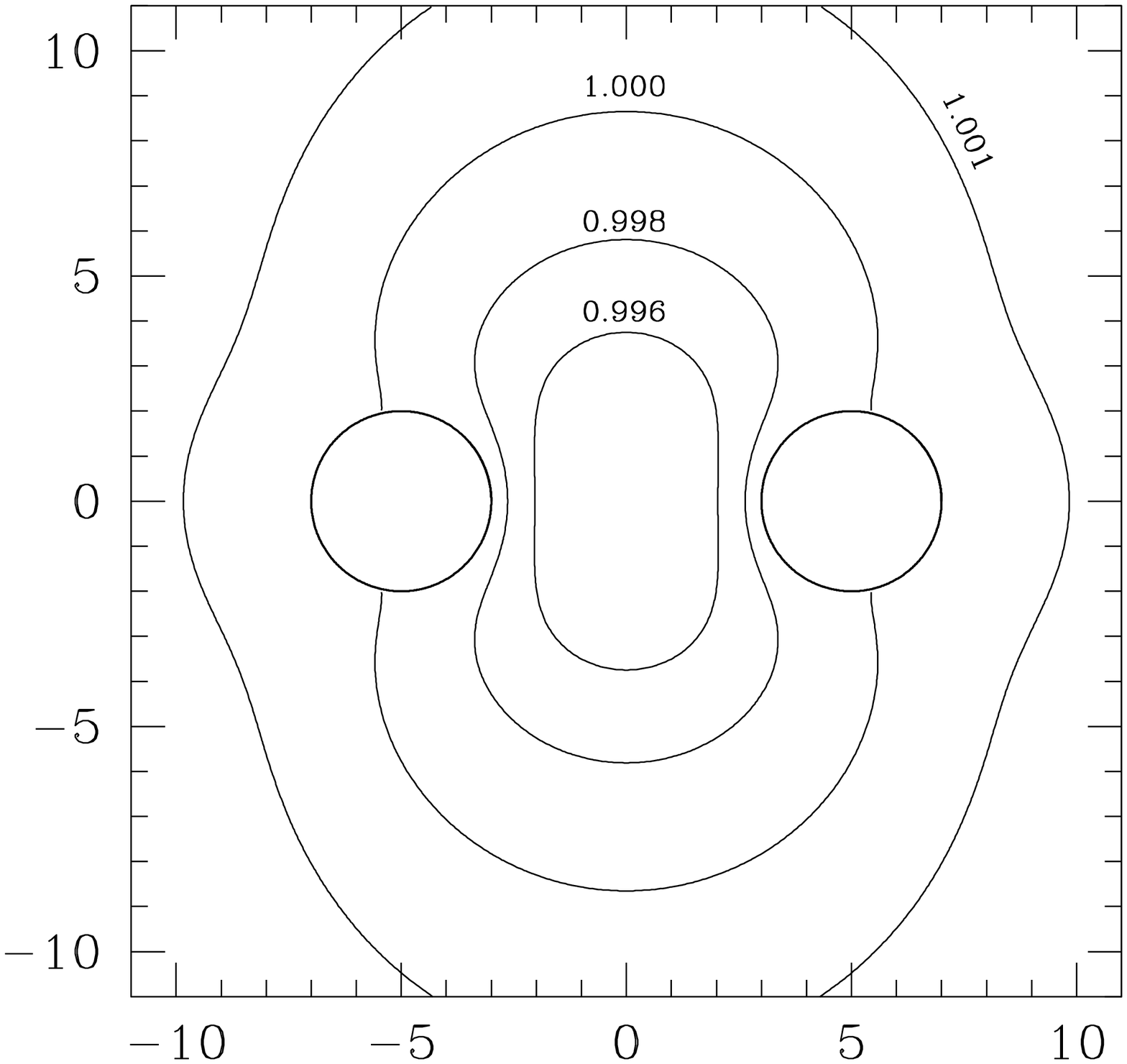}
\includegraphics[scale=0.23]{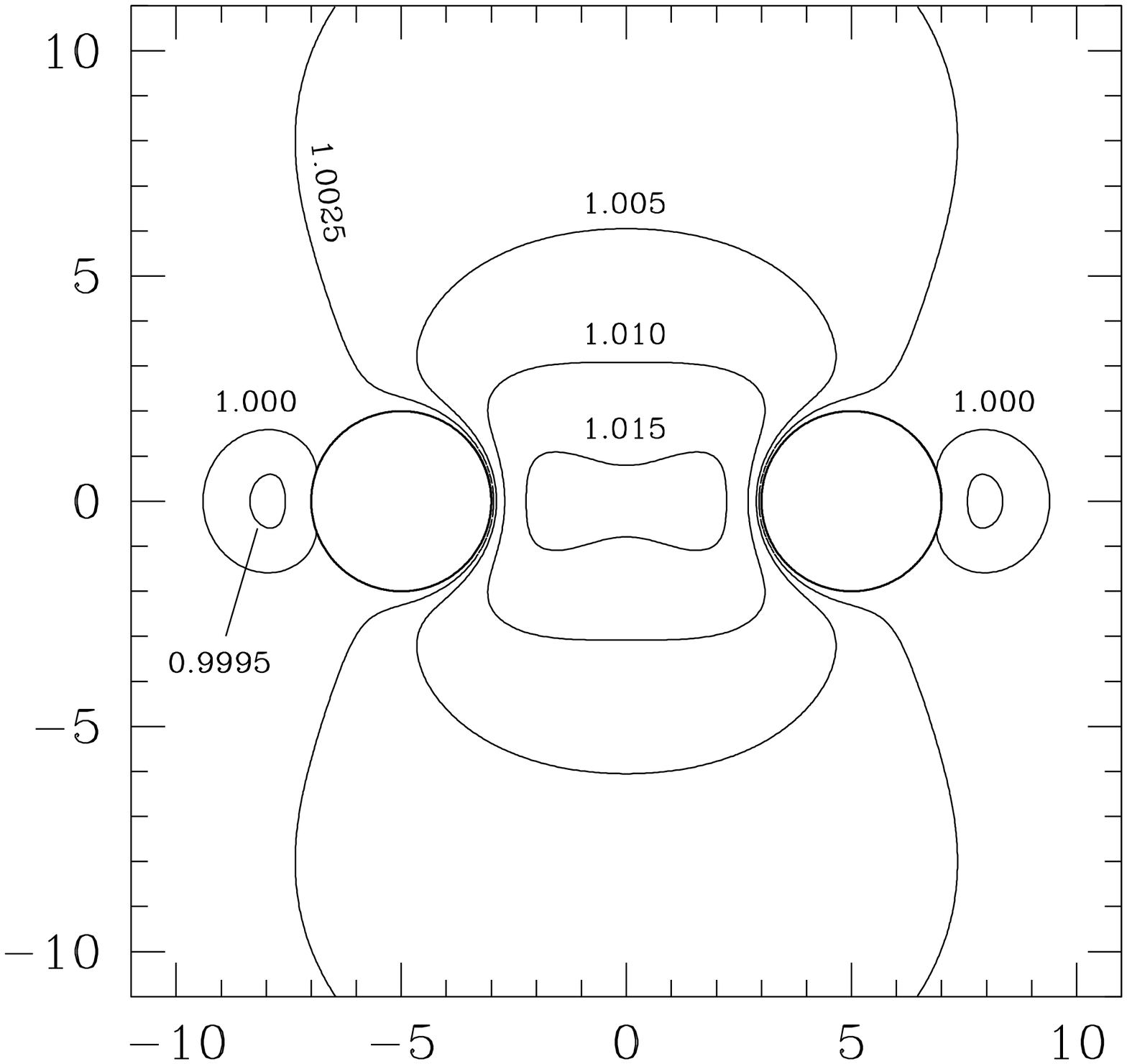}
}
\caption{\label{fig:Contours}
  Black holes at rest: Contour plots of the conformal factor $\psi$
  for ConfTT (left) and CTS-add (right). The circles denote the
  excised spheres of radius $2$. }
\end{figure}

Figure~\ref{fig:Compare-cuts} presents a plot of the conformal factor
along the axis through the centers of the holes. One sees that $\psi$
is close to $1$; however, between the holes ConfTT and CTS force
$\psi$ in {\em opposite} directions. For CTS, $\psi>1$ between the
holes, for ConfTT, $\psi<1$!  The contour plots in Fig.~\ref{fig:Contours} 
also show this striking difference between the
decompositions.

\begin{table*}
\caption{\label{tab:EADM}
  Solutions of different decompositions for two black holes at rest.
  Ham and Mom are  the rms residuals of the Hamiltonian 
  and momentum constraint, $\ell$ is the proper separation between 
  the apparent horizons. {\em mConfTT} represents the modified conformal TT
  decomposition which is explained in section \ref{sec:mConfTT}. 
  {\em inv. symm.} represents a conformally flat, time symmetric 
  and inversion symmetric  solution of the Hamiltonian constraint.
}
\begin{ruledtabular}
\begin{tabular}{ccccccccccc}
       & Ham & Mom & $E_{ADM}$ & $A_{AH}$ & $M_{AH}$ & $\ell$ & $\ell/m$ & $E_{ADM}/m$ & $E_{MPRC}/E_{ADM}$ & $E_b/\mu$ 
\rule[-.65em]{0.em}{1.3em}\\\hline
ConfTT & $9\times 10^{-7}$ & $4\times 10^{-7}$ & 2.06486 & 57.7369 & 1.07175 & 8.062 & 3.761 & 0.9633 & 0.2660 & -0.1467\\
PhysTT & $9\times 10^{-7}$ & $3\times 10^{-7}$ & 2.06490 & 57.7389 & 1.07176 & 8.062 & 3.761 & 0.9633 & 0.2660 & -0.1467\\
CTS-add  & $2\times 10^{-6}$ & $4\times 10^{-7}$ & 2.08121 & 62.3116 & 1.11340 & 8.039 & 3.610 & 0.9346 & 0.2434 & -0.2615\\ \hline
CTS-mult & $2\times 10^{-6}$ & $5\times 10^{-7}$ & 2.05851 & 60.8113 & 1.09991 & 8.080 & 3.672 & 0.9358 & 0.2444 & -0.2569\\
mConfTT & $3\times 10^{-6}$ & $1\times 10^{-6}$ & 2.0827 & 62.404 & 1.1142 &&& 0.9346 & 0.2434 & -0.2617\\ \hline
inv. symm. & - & -       & 4.36387 & 284.851 & 2.38053 & 17.731 & 3.724 & 0.9166 & 0.2285 & -0.3337
\end{tabular}
\end{ruledtabular}
\end{table*}

The result of PhysTT was found to be almost identical with ConfTT.
This is reasonable, since these two decompositions differ only in that
in one case the TT decomposition is with respect to the conformal
metric, and in the other case the TT decomposition it is with respect
to the physical metric. Since $\psi\approx 1$, the conformal metric is
almost identical to the physical metric, and only minor differences
arise. In the following we will often use ConfTT/PhysTT when referring
to both data sets.

We performed apparent horizon searches for these cases. For all
data sets, the apparent horizon is outside the sphere with radius
$2M$, that is outside the coordinate location for the apparent horizon
in a single hole spacetime.
For ConfTT/PhysTT, the radius of the apparent horizon surface is
$\approx 2.05M$, for CTS it is $\approx 2.15M$.  We computed
the apparent horizon area $A_{AH}$, the apparent horizon mass
\begin{equation}
  \label{eq:MAH}
  M_{AH}=\sqrt{\frac{A_{AH}}{16\pi}}
\end{equation}
of either hole, and the combined mass of both holes, 
\begin{equation}\label{eq:m}
  m=2M_{AH}.
\end{equation}
There is no rigorous definition of the mass of an individual black
hole in a binary black hole spacetime, and Eq.~(\ref{eq:MAH})
represents the true mass on an individual black hole
only in the limit of wide separation of the black holes.
A hard upper limit on the possible gravitational radiation
emitted to infinity during the coalescence process of a binary
will be
\begin{equation}\label{eq:MPRC}
  E_{MPRC}=E_{ADM}-\sqrt{\frac{2A_{AH}}{16\pi}},
\end{equation}
where $2A_{AH}$ is the combined apparent horizon area of both holes.
Thus, $E_{MPRC}$ represents the maximum possible radiation content
(MPRC) of the initial data.  This, of course, makes the unlikely
assumption that the binary radiates away all of its angular momentum.

We also compute the proper separation~$\ell$ between the apparent
horizon surfaces along the straight line connecting the centers of the excised
spheres.  In order to compare different data sets we consider the
dimensionless quantities $\ell/E_{ADM}$, $E_{ADM}/m$ and
$E_{MPRC}/E_{ADM}$.  We will also use $E_b/\mu$ which will be defined
shortly.

Table~\ref{tab:EADM} summarizes these quantities for all three
decompositions.  It also includes results for inversion symmetric
initial data, which for black holes at rest reduces to the Misner
data\cite{Misner:1963}\footnote{Although the Misner solution can be
  obtained analytically, we found it more convenient to solve the
  Hamiltonian constraint numerically.  The configuration in 
  Table~\ref{tab:EADM} corresponds to a separation $\beta=12$ in terms of
  \cite{Cook:1991}.}.  The results in Table~\ref{tab:EADM} are
intended to represent nearly the {\em same} physical configuration.

From Table~\ref{tab:EADM}, one finds that the black holes have roughly
the same dimensionless proper separation.  However, the scaled
ADM-energy $E_{ADM}/m$ differs by as much as 4.7\% between the
different data sets.  $E_{MPRC}/E_{ADM}$, which does not depend on any
notion of individual black hole masses at all, differs by 16\% between
the different data sets.

The inversion symmetric data has lowest $E_{ADM}/m$ and
$E_{MPRC}/E_{ADM}$, CTS has somewhat larger values, and ConfTT/PhysTT
lead to the biggest values for $E_{ADM}/m$ and $E_{MPRC}/E_{ADM}$.
This indicates that, relative to the sizes of the black holes,
ConfTT/PhysTT and CTS probably contain some excess energy.

A slightly different argument uses the binding energy which is defined
as
\begin{equation}
  \frac{E_b}{\mu}\equiv\frac{E_{ADM}-2M_{AH}}{\mu},
\end{equation}
where $\mu=M_{AH}/2$ is the reduced mass.  Two Newtonian point masses
at rest satisfy
\begin{equation}\label{eq:Eb-Newton}
  \frac{E_b}\mu = -\frac{1}{\ell/m}.
\end{equation}
From Table~\ref{tab:EADM} we see that for ConfTT/PhysTT, $|E_b/\mu| >
(l/m)^{-1}$, and for CTS, $|E_b/\mu|\approx (l/m)^{-1}$. Since gravity
in general relativity is typically {\em stronger} than Newtonian
gravity, we find again that CTS and ConfTT/PhysTT contain too much
energy relative to the black hole masses, ConfTT/PhysTT having even
more than CTS.

The proper separation between the apparent horizons $\ell/m$ is about
4\% smaller for CTS than for ConfTT/PhysTT. By
Eq.~(\ref{eq:Eb-Newton}) this should lead to a relative difference in binding
energy of the same order of magnitude.  Since $E_b/\mu$ differs by
almost a factor of two between the different decompositions, the
differences in $\ell/m$ play only a minor role.

\subsection{Configurations with angular momentum}
\label{sec:resultOrbiting}

Now we consider configurations which are approximating two black holes
in orbit around each other. The conformal metric is still a
superposition of two Kerr Schild metrics. The black holes are located
along the $x$-axis with a coordinate separation $d$.  For
ConfTT/PhysTT, we boost the individual holes to some velocity $\pm
v\hat e_y$ along the $y$-axis. For CTS we go to a co-rotating frame
with an angular frequency $\vec{\Omega}=\Omega\hat e_z$.  Thus, for
each decomposition we have a two parameter family of solutions, the
parameters being $(d,v)$ for ConfTT and PhysTT, and
$(d,\Omega)$ for CTS.

By symmetry, these configuration will have an ADM angular momentum
parallel to the $z$-axis which we denote by $J$. In order to compare
solutions among each other, and against the conformally flat inversion
symmetric data sets, we adjust the parameters $(d,v)$ and
$(d,\Omega)$, such that each initial-data set has angular momentum
$J/\mu m=2.976$ and a proper separation between the apparent horizons
of $l/m=4.880$. In Ref.~\cite{Cook:1994}, these values were determined to be
the angular momentum and proper separation of a binary black hole at
the innermost stable circular orbit.

\begin{table*}
\caption{\label{tab:EADM2}
Initial-data sets generated by different decompositions for binary 
black holes with the same angular momentum $J/\mu m$ and separation $\ell/m$.
The mConfTT dataset is explained in section \ref{sec:mConfTT}. It should 
be compared to CTS-add.
}
\begin{ruledtabular}
\begin{tabular}{ccccccccc}
 & parameters & $M_{AH}$ & $E_{ADM}$ & $J/\mu m$ & $\ell/m$ & $E_{ADM}/m$ & $E_{MPRC}/E_{ADM}$ & $E_b/\mu$
\rule[-.65em]{0.em}{1.3em}\\\hline
ConfTT & $d=11.899, v=0.26865$ & 1.06368 & 2.12035 & 2.9759 & 4.879 & 0.9967 & 0.2906 & -0.0132\\
PhysTT & $d=11.899, v=0.26865$ & 1.06369 & 2.12037 & 2.9757 & 4.879 & 0.9967 & 0.2906 & -0.0132\\
CTS-add  & $d=11.860, \Omega=0.0415$ & 1.07542 & 2.10391 & 2.9789 & 4.884 & 0.9782 & 0.2771 &-0.0873\\
CTS-mult & $d=11.750, \Omega=0.0421$ & 1.06528 & 2.08436 & 2.9776 & 4.880 & 0.9783 & 0.2772 & -0.0867\\
\hline
mConfTT & $d=11.860, \Omega=0.0415$ & 1.0758 & 2.1061 & 3.011 & 4.883 & 0.979 & 0.278 & -0.085\\
inv. symm.\footnote{Data taken from \cite{Cook:1994}} & & & & 2.976 & 4.880 & 0.9774 & 0.2766 & -0.09030
\end{tabular}
\end{ruledtabular}
\end{table*}

Table~\ref{tab:EADM2} lists the parameters corresponding to this
situation as well as results for each initial-data set\footnote{
  Because of the Lorentz contraction, the apparent horizons for
  ConfTT/PhysTT intersect the sphere with radius $2$. In order to have
  the full apparent horizon inside the computational domain, the
  radius of the excised spheres was reduced to 1.9 for these data
  sets.}.  As with the configuration with black holes at rest, we find
again that ConfTT/PhysTT and CTS lead to different ADM-energies.
Now, $E_{ADM}/m$ and $E_{MPRC}/E_{ADM}$ differ by $0.02$ and $0.013$,
respectively, between CTS and ConfTT/PhysTT.  However, in contrast to
the cases where the black holes at rest, now CTS and the inversion symmetric data set
have very similar values for $E_{ADM}/m$ and $E_{MPRC}/E_{ADM}$.

\subsection{Reconciling conformal TT and thin sandwich}
\label{sec:mConfTT}

We now investigate further the difference between ConfTT/PhysTT and
CTS.  Since the resulting initial-data sets for PhysTT and ConfTT are
very similar, we restrict our discussion to ConfTT.

\subsubsection{Motivation}

The construction of binary black hole data for the ConfTT/PhysTT
cases produces an extrinsic curvature that almost certainly
contains a significant TT component.  It would be interesting to know
how significant this component is to the value of the various physical
parameters we are comparing.  Ideally, we would like to completely
eliminate the TT component and see what effect this has on the resulting
data sets.  Unfortunately, this is a difficult, if not impossible,
task.

The TT component of a symmetric tensor $M^{ij}$ is defined as
\begin{equation}
        M^{ij}_{TT} \equiv M^{ij} - (\Long Y)^{ij},
\end{equation}
where the vector $Y^i$ is obtained by solving an elliptic equation
of the form
\begin{equation}\label{eq:Long_eqn}
        \LapLong Y^i = \nabla_jM^{ij}.
\end{equation}
The problem resides in the fact that the meaning of the TT component
depends of the boundary conditions used in solving (\ref{eq:Long_eqn}).

For the ConfTT/PhysTT cases we are actually solving for a vector $V^i$
that is a linear combination of two components, one that solves an
equation of the form of (\ref{eq:Long_eqn}) to obtain the TT component
of $\tilde{M}^{ij}$ and one that solves the momentum constraint.  But
by imposing inner-boundary conditions of $V^i=0$, we don't specify the
boundary conditions on either part independently.  Nor is it clear
what these boundary conditions should be.  Since we cannot explicitly
construct the TT component of the extrinsic curvature, we cannot
eliminate it.  Although it is not ideal, there is an alternative we
can consider that does provide some insight into the importance of
the initial choice of $\tilde{M}^{ij}$.

\subsubsection{Black holes at rest}
\label{sec:mConfTT:Rest}

Consider the following numerical experiment for two black holes at
rest: Given $\tilde M^{ij}$ from Eq.~(\ref{eq:BinaryKerrSchild-Mij}),
make a transverse traceless decomposition by setting
\begin{equation}\label{eq:mConfTT}
  2N \tilde M^{ij}=\tilde M_{TT}^{ij}+(\tilde\Long Y)^{ij}
\end{equation}
where $\tilde\nabla_j \tilde M^{ij}_{TT}=0$ and $N=N_A+N_B-1$.  
Notice that we are decomposing $2N\tilde{M}^{ij}$, not $\tilde{M}^{ij}$.
Taking
the divergence of Eq.~(\ref{eq:mConfTT}) one finds 
\begin{equation}\label{eq:mConfTT-X}
  \tildeLapLong Y^i=\tilde\nabla_j\left(2N\tilde M^{ij}\right).
\end{equation}
The decomposition chosen in Eq.~(\ref{eq:mConfTT}) is motivated by the
conformal thin sandwich decomposition. With this decomposition we can,
in fact, use the shift vector $N^i$ to fix boundary conditions on
$Y^i$, just as it was used to fix the boundary conditions in
Eqs.~(\ref{eq:Sandwich-BC2}---\ref{eq:Sandwich-BC4}).  For the black
holes at rest in this case, we have $\Omega=0$. After solving
Eq.~(\ref{eq:mConfTT-X}) for $Y^i$, we can construct a new 
conformal extrinsic curvature by
\begin{equation}\label{eq:mConfTT-Mij}
  {\tilde{M}{'}}^{\;ij}=\frac{1}{2N}(\tilde\Long Y)^{ij}
\end{equation}
which is similar to what would result if we could eliminate $\tilde
M^{ij}_{TT}$ from $\tilde M^{ij}$. Using $\tilde M^{\prime ij}$ in
place of $\tilde M^{ij}$, we can again solve the conformal TT
equations.  The result of this modified conformal TT decomposition
{\bf ``mConfTT''} is striking: Figure~\ref{fig:Compare-cuts} shows
that mConfTT generates a conformal factor $\psi$ that is very similar
to $\psi$ of CTS. mConfTT is also included in Table~\ref{tab:EADM}
where it can be seen that the quantities $E_{ADM}/m$ and
$E_{MPRC}/E_{ADM}$ differ only slightly between mConfTT and CTS.

The fact that modification of the extrinsic curvature changes the
ADM-energy by such a large amount underlines the importance of a
careful choice for the extrinsic curvature $\tilde M^{ij}$ in
ConfTT/PhysTT.  The extremely good agreement between CTS and mConfTT
is probably caused by our procedure to determine $\tilde M{'}^{ij}$.
We force the extrinsic curvature of mConfTT into the form
Eq.~(\ref{eq:mConfTT-Mij}). This is precisely the form of the
extrinsic curvature in CTS, Eq.~(\ref{eq:SandwichTT-3}), even using
the same function $N$ and the same boundary conditions on the vectors
$Y^i$ and $\beta^i$.

\subsubsection{Black holes with angular momentum}

We now apply the modified conformal TT decomposition to the orbiting
configurations of section~\ref{sec:resultOrbiting}.  In the corotating
frame, the black holes are at rest, and we start with
$\tilde\gamma_{ij}$ and $\tilde M^{ij}$ of two black holes {\em at
  rest} with coordinate separation $d=11.860$. We now solve
Eq.~(\ref{eq:mConfTT-X}) with
\begin{equation}
  N=N_A+N_B-1
\end{equation}
and corotating boundary conditions on $Y^i$ [cf.\
Eqs.~(\ref{eq:Sandwich-BC2})--(\ref{eq:Sandwich-BC4})]:
\begin{subequations}
\begin{align}
Y^i&=N_{\!A}^i                 &&\mbox{sphere inside hole A,}\\
Y^i&=N_B^i                     &&\mbox{sphere inside hole B,}\\
Y^i&=\vec\Omega\times\vec r    &&\mbox{outer boundary.}
\end{align}
\end{subequations}
$N_{A/B}$ and $N_{A/B}^i$ are lapse and shift of individual Kerr-Schild
black holes at rest.  $\tilde M{'}^{ij}$ is again constructed by
Eq.~(\ref{eq:mConfTT-Mij}) and used in solving the conformal TT equations.

Results from this procedure are included in Table~\ref{tab:EADM2}.
Again, mConfTT generates results very close to CTS.  $E_{ADM}/m$ 
changes by 1.8\% of the total mass between ConfTT and mConfTT, again
highlighting the importance of the extrinsic curvature.

\subsection{Dependence on the size of the excised spheres}

The framework presented in this paper requires the excision of the
singularities at the centers of each hole\footnote{Marronetti and
  Matzner\cite{Marronetti-Matzner:2000} effectively excised the
  centers, too, by using ``blending functions''.}.  So far we have used
$r_{exc}=2M$ or $r_{exc}=1.9M$ in order to impose boundary conditions
close to the apparent horizons, but different choices can be made.
Indeed, one might expect that the boundary conditions
(\ref{eq:BC-ConfTT-PhysTT}) and (\ref{eq:BC-sandwich}) become ``better''
farther inside the apparent horizon, where the metric and extrinsic
curvature of that black hole dominate the superposed metric
$\tilde\gamma_{ij}$ and superposed extrinsic curvature $\tilde
M^{ij}$.

In order to test this assumption, we solve the constraint equations
for two black holes at rest for different radii $r_{exc}$. We find
that for all three decompositions, the data sets depend strongly on
the radius of the excised spheres.

\begin{figure}
  \centering
  \includegraphics[scale=0.38]{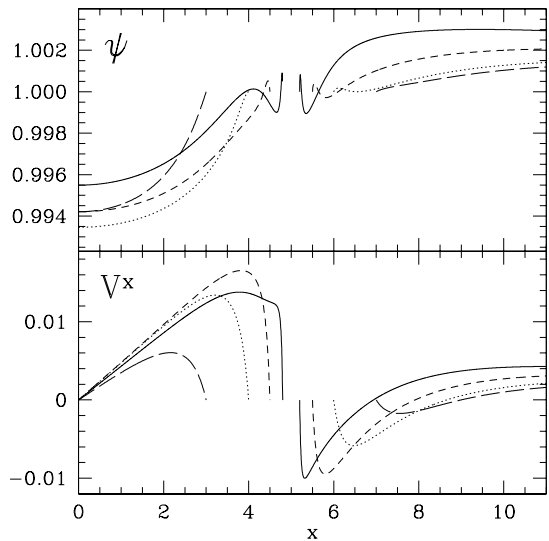}
  \caption{Plots of $\psi$ and $V^x$ along the positive $x$-axis
    for ConfTT for different radii $r_{exc}=2M, M, 0.5M, 0.2M$. The
    excised spheres are centered on the $x$-axis at $x=\pm 5$.  The
    position where a line terminates gives $r_{exc}$ for that line.}
    \label{fig:ConfTT-cuts}
 \end{figure}

\begin{figure}
  \centerline{\includegraphics[scale=0.38]{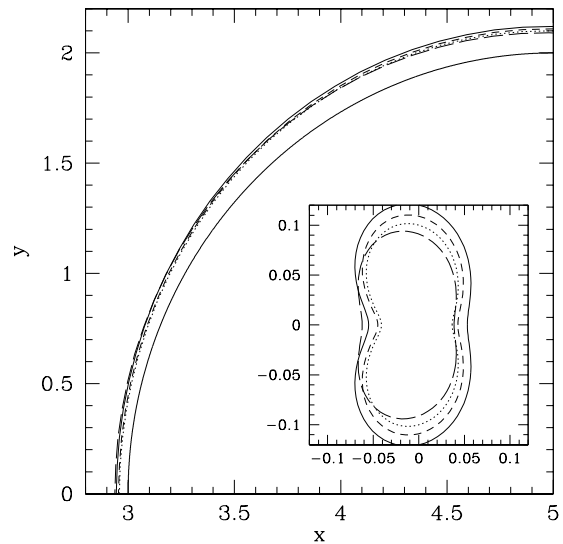}}
  \caption{\label{fig:ConfTT-AH}Apparent horizons for ConfTT
    with different radii of excised spheres. Results shown are for
    $r_{exc}=2M$ (long dashed line), $M$ (dotted line), $0.5M$ (short
    dashed line) and $0.2M$ (outer solid line). The inner solid line
    is a circle with radius 2.  The insert shows a parametric plot of
    $r(\phi)-2$, which emphasizes the differences between the
    different apparent horizons.}
\end{figure}

\begin{table}
\caption{\label{tab:TT}
Solutions of solving ConfTT for different radii of
the excised spheres, $r_{exc}$.  The results for PhysTT
are nearly identical.
}
\begin{ruledtabular}
\begin{tabular}{cccccc}
$r_{exc}$ & $E_{ADM}$ & $A_{AH}$ & $\ell$ & $E_{ADM}/m$ & $\ell/E_{ADM}$
\rule[-.65em]{0.em}{1.3em}\\\hline
\multicolumn{6}{c}{ Conformal TT }\\
2.0 &    2.0649 &    57.737 &   8.062 &   0.9633 &   3.904 \\ 
1.0 &    2.0682 &    57.825 &   8.101 &   0.9641 &   3.917 \\ 
0.5 &    2.0808 &    58.520 &   8.101 &   0.9642 &   3.893 \\ 
0.2 &    2.0978 &    59.514 &   8.093 &   0.9640 &   3.858 \\ 
0.1 &    2.1064 &    60.025 &   8.089 &   0.9638 &   3.840 \\ 
\end{tabular}
\end{ruledtabular}
\end{table}

Figure~\ref{fig:ConfTT-cuts} presents plots of the conformal factor
$\psi$ and $V^x$ for ConfTT with different $r_{exc}$.  There is no
clear sign of convergence of $\psi$ as $r_{exc}\to 0$.  For
$r_{exc}=0.2M$, the conformal factor $\psi$ even oscillates close to
the excised sphere.  Table~\ref{tab:TT} displays various
quantities for the ConfTT decomposition for
different $r_{exc}$.
As $r_{exc}$ varies
between $2.0M$ and $0.1M$, the ADM-energy varies between $2.065$ and
$2.106$, whereas the apparent horizon area changes by nearly 4\%.  The
apparent horizons move around somewhat as $r_{exc}$ changes. 
Figure~\ref{fig:ConfTT-AH} shows the location of the apparent horizons for
different $r_{exc}$.

\begin{figure}
  \centering
  \includegraphics[scale=0.38]{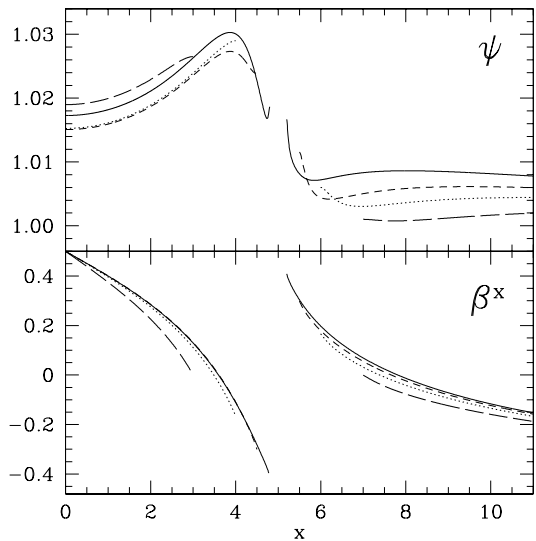}
  \caption{\label{fig:Sandwich-cuts-N1N2} 
    Cuts through $\psi$ and $\beta^x$ for CTS-mult for different radii
    $r_{exc}$. Here $\tilde \alpha=N_{\!A}N_B$ and the boundary
    condition on $\psi$ at the excised spheres is $d\psi/dr=0$.  The
    curves for $\beta^x$ are shifted up by 0.5 for $x<5$, and are
    shifted down by 0.5 for $x>5$ to allow for better plotting.
    $d\psi/dr$ approaches zero at the inner boundary on scales too
    small to be seen in this figure.  
}  
\end{figure}

\begin{figure}
 \centerline{\includegraphics[scale=0.38]{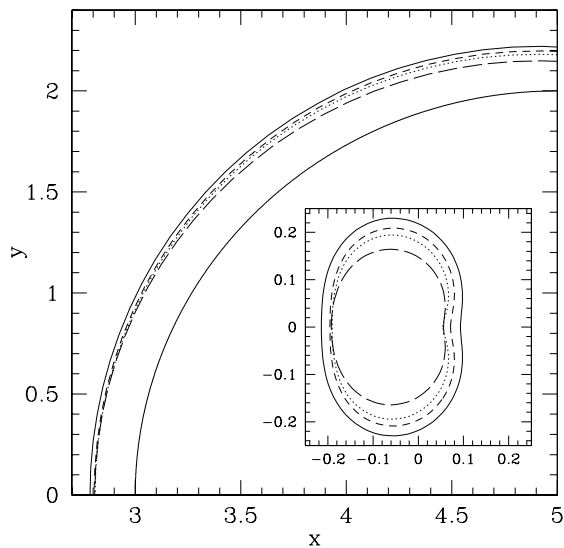}}
 \caption{\label{fig:Sandwich-AH-2} 
   Apparent horizons for CTS with $\tilde\alpha=N_A\,N_B$ and inner
   boundary condition $d\psi/dr=0$.  The different curves belong to
   different $r_{exc}$ as explained in Fig.~\ref{fig:ConfTT-AH} }
\end{figure}

For CTS-add (with $\tilde\alpha=N_A+N_B-1$), the initial-data sets
seem to diverge as $r_{exc}$ is decreased.  This has to be expected,
since this choice for $\tilde\alpha$ changes sign if the excised
spheres become sufficiently small.  Changing to $\tilde\alpha=N_AN_B$
so that the lapse does not change sign reduces this divergent behavior.  Von
Neumann boundary conditions on $\psi$ at the excised spheres,
\begin{equation}
  \frac{\partial\psi}{\partial r}=0,
\end{equation}
lead to an increase in $A_{AH}$ especially for large excised spheres.
This combination of lapse $\tilde\alpha$ and boundary conditions
exhibits the smallest variations in $E_{ADM}/m$; cuts through $\psi$,
$\beta^x$ and through the apparent horizons are shown in
Figs.~\ref{fig:Sandwich-cuts-N1N2} and \ref{fig:Sandwich-AH-2}.  From
the three examined combinations of lapse and boundary conditions, the
one shown behaves best, but there is still no convincing sign of
convergence.

Table~\ref{tab:CTS} presents ADM-energies and apparent
horizon areas and masses for CTS with different $r_{exc}$ and
different choices of lapse and boundary condition.  From the unscaled
ADM-energy $E_{ADM}$ it is apparent that $\tilde\alpha=N_A+N_B-1$
diverges most strongly.  Note that between all choices of lapse,
boundary conditions and $r_{exc}$, the unscaled quantities $E_{ADM}$,
$M_{AH}$, and $\ell$ exhibit a much broader variation than the scaled
quantities $E_{ADM}/m$ and $\ell/E_{ADM}$.

\begin{table}
\caption{\label{tab:CTS}
  Solutions of CTS as a function of radius of excised
  spheres, $r_{exc}$.  Different choices of the lapse $\tilde\alpha$ and
  boundary conditions for $\psi$ at the excised spheres are explored.
}
\begin{ruledtabular}
\begin{tabular}{ccrccc}
$r_{exc}$ & $E_{ADM}$ & $A_{AH}\;$& $\ell$ & $E_{ADM}/m$ & $\ell/E_{ADM}$ 
\rule[-.5em]{0.em}{1.5em}\\\hline
\multicolumn{6}{c}{ $\tilde\alpha=N_A+N_B-1,\;\; \psi=1$ }\\
2.0 &    2.0812 &    62.312 &   8.039 &   0.9346 &   3.863 \\ 
1.0 &    2.1846 &    68.279 &   8.000 &   0.9372 &   3.662 \\ 
0.5 &    2.3085 &    76.253 &   7.925 &   0.9371 &   3.433 \\ 
0.2 &    2.5463 &    93.534 &   7.750 &   0.9333 &   3.044 \\ 
0.1 &    2.8543 &   118.834 &   7.489 &   0.9282 &   2.624 \\ 
\hline
\multicolumn{6}{c}{ $\tilde\alpha=N_AN_B,\;\; \psi=1$ }\\
2.0 &    2.0585 &    60.811 &   8.080 &   0.9358 &   3.925 \\ 
1.0 &    2.1216 &    64.080 &   8.044 &   0.9395 &   3.792 \\ 
0.5 &    2.1696 &    66.790 &   8.017 &   0.9411 &   3.695 \\ 
0.2 &    2.2120 &    69.456 &   7.991 &   0.9409 &   3.613 \\ 
0.1 &    2.2326 &    70.809 &   7.978 &   0.9405 &   3.573 \\ 
\hline
\multicolumn{6}{c}{ $\tilde\alpha=N_AN_B,\;\; \partial\psi/\partial r=0$ }\\
2.0 &    2.1110 &    64.229 &   8.085 &   0.9337 &   3.830 \\ 
1.0 &    2.1533 &    66.128 &   8.030 &   0.9387 &   3.729 \\ 
0.5 &    2.1794 &    67.427 &   8.011 &   0.9409 &   3.676 \\ 
0.2 &    2.2136 &    69.559 &   7.990 &   0.9409 &   3.609 \\ 
0.1 &    2.2330 &    70.836 &   7.978 &   0.9405 &   3.573 \\ 
\end{tabular}
\end{ruledtabular}
\end{table}

\section{Discussion}
\label{sec:Discussion}

Our results clearly show that different decompositions lead to
different initial-data sets, even when seemingly similar
choices for the freely specifiable pieces are used.  From
Tables~\ref{tab:EADM} and \ref{tab:EADM2}, one sees that $E_{ADM}/m$
changes by as much as 0.029 between ConfTT/PhysTT and CTS.  The
difference between ConfTT/PhysTT and the inversion symmetric data is
even larger, 0.047.  These numbers seem to be small; however, current
evolutions of binary data usually find the total energy emitted in
gravitational radiation $E_{GW}/m$ to be between 0.01 and 0.03
\cite{Baker-Bruegmann-etal:2001,Alcubierre:2001,Brandt-Corell-etal:2000},
which is the same order of magnitude as the changes in $E_{ADM}/m$ we
find.  This means that, in principle, most of the energy radiated in
these simulations could originate from ``spurious'' energy in the
system and not from the dynamics of the binary system we are
interested in.

These findings highlight the fact that current binary black hole
initial data sets are inadequate for the task of accurately describing
realistic binary systems.  We see that the choices of the conformal
3-geometry $\tilde\gamma_{ij}$ and the freely specifiable portions of
the extrinsic curvature, embedded in $\tilde{M}^{ij}$, influence the
content of the initial data at a significant level.  Furthermore, the
results suggest that small changes in the free data associated with
the extrinsic curvature are more significant than small changes in the
choice of $\tilde\gamma_{ij}$.\footnote{Following submission of this
paper, a preprint by Damour et
al.\cite{Damour-Gourgoulhon-Grandclement:grqc0204011} has appeared
that lends support to our idea that the extrinsic curvature plays a
key role in constructing quasi-equilibrium binary black hole initial
data.}  This assertion is supported by the fact that $E_{ADM}/m$
is consistently larger for the ConfTT solutions than for the CTS
solutions but the two approaches can be made to produce quite
consistent results by using the modified extrinsic curvature of the
mConfTT method.  All of these decompositions use the same non-flat
conformal metric, but differ in the extrinsic curvature. On the other
hand, results for the conformally-flat inversion-symmetric data agree
rather well with the results from the CTS method when we consider
orbiting black holes.  For black holes at rest, CTS differs from the
inversion symmetric data, which seems to contradict our conclusion.
However, this difference is likely due to the time-symmetric nature of
the inversion symmetric data, which is especially adapted to the
time-symmetry of the particular configuration of ``two black holes at
rest''.

Improved binary black hole initial data will require choices for the
freely specifiable data that are physically motivated, rather than
chosen for computational convenience.  The same is true for the
boundary conditions used in solving the constraints.  The boundary
conditions used in this paper carry the implicit assumptions that the
approximate metric and extrinsic curvature are correct at the
excision boundaries and that the value of the single-hole Kerr-Schild
shift at the excision boundary is correct in a multi-hole situation.
This is clearly not true, but we might hope that the impact of the
error in this choice would diminish as we decrease the radius of the
excision boundary.  However, our results presented in
Tables~\ref{tab:TT} and \ref{tab:CTS} do not support this conjecture.
Examining the change in $E_{ADM}/m$ as we vary $r_{exc}$ shows only a
small change, but more importantly, it shows no sign of converging as
we decrease $r_{exc}$.  The effects of changing $r_{exc}$ are much
more significant for $\ell/E_{ADM}$, changing its value by as much as
10\% in the case of CTS-mult for the range of values considered.
Furthermore, as with the energy, we see no sign of convergence in
$\ell/E_{ADM}$ as $r_{exc}$ decreases.  Interestingly, although the
solutions show no sign of convergence as we shrink the excision
radius, we do find that the dimensionless quantities $E_{ADM}/m$ and
$\ell/E_{ADM}$ do become independent of the choice of the
inner-boundary condition as $r_{exc}$ decreases.  This can be seen in
comparing the result in Table~\ref{tab:CTS} for the cases using
$\psi=1$ and $\partial\psi/\partial r=0$ as inner-boundary conditions.
Additional tests, not reported in this paper, further support this
assertion.

\section{Conclusion}

Using a new elliptic solver capable of solving the initial-value
problem of general relativity for any of three different
decompositions and any choice for the freely specifiable data, we have
examined data sets representing binary black hole spacetimes.  We find
that the choices for the freely specifiable data currently in use are
inadequate for the task of simulating the gravitational radiation
produced in astrophysically realistic situations.  In particular,
we studied the results of using a superposition of two Kerr-Schild
black holes to fix the freely specifiable data and compared them
to the results obtained from conformally flat initial data.

Although the new Kerr-Schild based data provide a valuable point of
comparison, it is not clear that the data produced are significantly
superior to previous conformally-flat data.  What is clear is that the
choice of the freely specifiable data will be very important in
constructing astrophysically realistic binary black hole initial data.
Progress will require that these data, {\em and} the boundary
conditions needed to solve the constraints, must be chosen based on
physical grounds rather than computational convenience.

How can better initial data be achieved and how can the quality of
initial data be measured?  We believe that the conformal thin sandwich
decomposition will be especially useful.  Genuine radiative degrees of
freedom cannot {\em in principle} be recognized on a single time
slice. The conformal thin sandwich method uses in effect two nearby
surfaces, giving it a potential advantage over other methods.  Also,
it avoids much of the uncertainty related to specifying a conformal
extrinsic curvature.
Moreover, the conformal thin sandwich approach is especially well
suited for the most interesting configurations, a black-hole binary in
a quasi-equilibrium orbit.  In this case time derivatives of all
quantities are small and the choice $\tilde u^{ij}=0$ is physically
motivated.  One should exploit the condition of quasi-equilibrium as
fully as possible, i.e. one should use the conformal thin sandwich
approach together with the constant $K$ equation, $\partial_tK=0$. The
latter yields another elliptic equation for the lapse which removes
the arbitrariness inherent in choosing a conformal lapse
$\tilde\alpha$.  One will also need more physical boundary conditions.
Work in this direction was begun
in\cite{Gourgoulhon-Grandclement-Bonazzola:2001a,
Grandclement-Gourgoulhon-Bonazzola:2001b} and refined in
\cite{Cook:2001}.

Ultimately, the gravitational wave content of an initial-data set can
be determined only by long term evolutions. One must compute an
initial-data set representing a binary black hole in quasi-circular
orbit and evolve it. Then one must repeat this process with an 
initial-data set representing the {\em same} binary black hole, say, one
orbital period earlier, and evolve that data set, too.  If both
evolutions lead to the same gravitational waves (modulo time offset)
then one can be confident that the gravitational radiation is indeed
astrophysically realistic.  This approach has recently been used for
the first time in conjunction with conformally flat puncture data,
where it proved remarkably successful
\cite{Baker-Campanelli-etal:astroph0202469}.

\acknowledgments We thank Lawrence Kidder, Mark Scheel, and James
York for helpful discussions.  This work was supported in part by NSF
grants PHY-9800737 and PHY-9900672 to Cornell University, and by NSF
grant PHY-9988581 to Wake Forest University.  Computations were
performed on the IBM SP2 of the Department of Physics, Wake Forest
University, with support from an IBM SUR grant.


\begin{thebibliography}{40}
\expandafter\ifx\csname natexlab\endcsname\relax\def\natexlab#1{#1}\fi
\expandafter\ifx\csname bibnamefont\endcsname\relax
  \def\bibnamefont#1{#1}\fi
\expandafter\ifx\csname bibfnamefont\endcsname\relax
  \def\bibfnamefont#1{#1}\fi
\expandafter\ifx\csname citenamefont\endcsname\relax
  \def\citenamefont#1{#1}\fi
\expandafter\ifx\csname url\endcsname\relax
  \def\url#1{\texttt{#1}}\fi
\expandafter\ifx\csname urlprefix\endcsname\relax\def\urlprefix{URL }\fi
\providecommand{\bibinfo}[2]{#2}
\providecommand{\eprint}[2][]{\url{#2}}

\bibitem[{\citenamefont{Baker et~al.}(2001)\citenamefont{Baker, Br{\"u}gmann,
  Campanelli, Lousto, and Takahashi}}]{Baker-Bruegmann-etal:2001}
\bibinfo{author}{\bibfnamefont{J.}~\bibnamefont{Baker}},
  \bibinfo{author}{\bibfnamefont{B.}~\bibnamefont{Br{\"u}gmann}},
  \bibinfo{author}{\bibfnamefont{M.}~\bibnamefont{Campanelli}},
  \bibinfo{author}{\bibfnamefont{C.~O.} \bibnamefont{Lousto}},
  \bibnamefont{and}
  \bibinfo{author}{\bibfnamefont{R.}~\bibnamefont{Takahashi}},
  \bibinfo{journal}{Phys. Rev. Lett.} \textbf{\bibinfo{volume}{87}},
  \bibinfo{pages}{121103} (\bibinfo{year}{2001}).

\bibitem[{\citenamefont{Alcubierre et~al.}(2001)\citenamefont{Alcubierre,
  Benger, Br{\"u}gmann, Lanfermann, Nerger, Seidel, and
  Takahashi}}]{Alcubierre:2001}
\bibinfo{author}{\bibfnamefont{M.}~\bibnamefont{Alcubierre}},
  \bibinfo{author}{\bibfnamefont{W.}~\bibnamefont{Benger}},
  \bibinfo{author}{\bibfnamefont{B.}~\bibnamefont{Br{\"u}gmann}},
  \bibinfo{author}{\bibfnamefont{G.}~\bibnamefont{Lanfermann}},
  \bibinfo{author}{\bibfnamefont{L.}~\bibnamefont{Nerger}},
  \bibinfo{author}{\bibfnamefont{E.}~\bibnamefont{Seidel}}, \bibnamefont{and}
  \bibinfo{author}{\bibfnamefont{R.}~\bibnamefont{Takahashi}},
  \bibinfo{journal}{Phys. Rev. Lett.} \textbf{\bibinfo{volume}{87}},
  \bibinfo{pages}{271103} (\bibinfo{year}{2001}).

\bibitem[{\citenamefont{Baker et~al.}(2002)\citenamefont{Baker, Campanelli,
  Lousto, and Takahashi}}]{Baker-Campanelli-etal:astroph0202469}
\bibinfo{author}{\bibfnamefont{J.}~\bibnamefont{Baker}},
  \bibinfo{author}{\bibfnamefont{M.}~\bibnamefont{Campanelli}},
  \bibinfo{author}{\bibfnamefont{C.~O.} \bibnamefont{Lousto}},
  \bibnamefont{and}
  \bibinfo{author}{\bibfnamefont{R.}~\bibnamefont{Takahashi}},
  \bibinfo{journal}{astro-ph/0202469}  (\bibinfo{year}{2002}).

\bibitem[{\citenamefont{York{,}~Jr.}(1979)}]{York:1979}
\bibinfo{author}{\bibfnamefont{J.~W.} \bibnamefont{York{,}~Jr.}}, in
  \emph{\bibinfo{booktitle}{Sources of Gravitational Radiation}}, edited by
  \bibinfo{editor}{\bibfnamefont{L.~L.} \bibnamefont{Smarr}}
  (\bibinfo{address}{Cambridge University Press, Cambridge, UK},
  \bibinfo{year}{1979}), pp. \bibinfo{pages}{83--126}.

\bibitem[{\citenamefont{Murchadha and
  York{,}~Jr.}(1974{\natexlab{a}})}]{Murchadha-York:1974a}
\bibinfo{author}{\bibfnamefont{N.~{\'O}.} \bibnamefont{Murchadha}}
  \bibnamefont{and} \bibinfo{author}{\bibfnamefont{J.~W.}
  \bibnamefont{York{,}~Jr.}}, \bibinfo{journal}{Phys. Rev. D}
  \textbf{\bibinfo{volume}{10}}, \bibinfo{pages}{428}
  (\bibinfo{year}{1974}{\natexlab{a}}).

\bibitem[{\citenamefont{Murchadha and
  York{,}~Jr.}(1974{\natexlab{b}})}]{Murchadha-York:1974b}
\bibinfo{author}{\bibfnamefont{N.~{\'O}.} \bibnamefont{Murchadha}}
  \bibnamefont{and} \bibinfo{author}{\bibfnamefont{J.~W.}
  \bibnamefont{York{,}~Jr.}}, \bibinfo{journal}{Phys. Rev. D}
  \textbf{\bibinfo{volume}{10}}, \bibinfo{pages}{437}
  (\bibinfo{year}{1974}{\natexlab{b}}).

\bibitem[{\citenamefont{Murchadha and York{,}~Jr.}(1976)}]{Murchadha-York:1976}
\bibinfo{author}{\bibfnamefont{N.~{\'O}.} \bibnamefont{Murchadha}}
  \bibnamefont{and} \bibinfo{author}{\bibfnamefont{J.~W.}
  \bibnamefont{York{,}~Jr.}}, \bibinfo{journal}{Gen. Relativ. Gravit.}
  \textbf{\bibinfo{volume}{7}}, \bibinfo{pages}{257} (\bibinfo{year}{1976}).

\bibitem[{\citenamefont{York{,}~Jr.}(1999)}]{York:1999}
\bibinfo{author}{\bibfnamefont{J.~W.} \bibnamefont{York{,}~Jr.}},
  \bibinfo{journal}{Phys. Rev. Lett.} \textbf{\bibinfo{volume}{82}},
  \bibinfo{pages}{1350} (\bibinfo{year}{1999}).

\bibitem[{\citenamefont{Bowen}(1979)}]{Bowen:1979}
\bibinfo{author}{\bibfnamefont{J.~M.} \bibnamefont{Bowen}},
  \bibinfo{journal}{Gen. Relativ. Gravit.} \textbf{\bibinfo{volume}{11}},
  \bibinfo{pages}{227} (\bibinfo{year}{1979}).

\bibitem[{\citenamefont{Bowen and York{,}~Jr.}(1980)}]{Bowen-York:1980}
\bibinfo{author}{\bibfnamefont{J.~M.} \bibnamefont{Bowen}} \bibnamefont{and}
  \bibinfo{author}{\bibfnamefont{J.~W.} \bibnamefont{York{,}~Jr.}},
  \bibinfo{journal}{Phys. Rev. D} \textbf{\bibinfo{volume}{21}},
  \bibinfo{pages}{2047} (\bibinfo{year}{1980}).

\bibitem[{\citenamefont{Kulkarni et~al.}(1983)\citenamefont{Kulkarni, Shepley,
  and York{,}~Jr.}}]{Kulkarni-Shepley-York:1983}
\bibinfo{author}{\bibfnamefont{A.~D.} \bibnamefont{Kulkarni}},
  \bibinfo{author}{\bibfnamefont{L.~C.} \bibnamefont{Shepley}},
  \bibnamefont{and} \bibinfo{author}{\bibfnamefont{J.~W.}
  \bibnamefont{York{,}~Jr.}}, \bibinfo{journal}{Phys. Lett. A}
  \textbf{\bibinfo{volume}{96A}}, \bibinfo{pages}{228} (\bibinfo{year}{1983}).

\bibitem[{\citenamefont{Thornburg}(1987)}]{Thornburg:1987}
\bibinfo{author}{\bibfnamefont{J.}~\bibnamefont{Thornburg}},
  \bibinfo{journal}{Class. Quantum Gravit.} \textbf{\bibinfo{volume}{4}},
  \bibinfo{pages}{1119} (\bibinfo{year}{1987}).

\bibitem[{\citenamefont{Cook et~al.}(1993)\citenamefont{Cook, Choptuik, Dubal,
  Klasky, Matzner, and Oliveira}}]{Cook-Choptuik-etal:1993}
\bibinfo{author}{\bibfnamefont{G.~B.} \bibnamefont{Cook}},
  \bibinfo{author}{\bibfnamefont{M.~W.} \bibnamefont{Choptuik}},
  \bibinfo{author}{\bibfnamefont{M.~R.} \bibnamefont{Dubal}},
  \bibinfo{author}{\bibfnamefont{S.}~\bibnamefont{Klasky}},
  \bibinfo{author}{\bibfnamefont{R.~A.} \bibnamefont{Matzner}},
  \bibnamefont{and} \bibinfo{author}{\bibfnamefont{S.~R.}
  \bibnamefont{Oliveira}}, \bibinfo{journal}{Phys. Rev. D}
  \textbf{\bibinfo{volume}{47}}, \bibinfo{pages}{1471} (\bibinfo{year}{1993}).

\bibitem[{\citenamefont{Brandt and Br{\"u}gmann}(1997)}]{Brandt-Brugmann:1997}
\bibinfo{author}{\bibfnamefont{S.}~\bibnamefont{Brandt}} \bibnamefont{and}
  \bibinfo{author}{\bibfnamefont{B.}~\bibnamefont{Br{\"u}gmann}},
  \bibinfo{journal}{Phys. Rev. Lett.} \textbf{\bibinfo{volume}{78}},
  \bibinfo{pages}{3606} (\bibinfo{year}{1997}).

\bibitem[{\citenamefont{Rieth}(1997)}]{Rieth:1997}
\bibinfo{author}{\bibfnamefont{R.}~\bibnamefont{Rieth}}, in
  \emph{\bibinfo{booktitle}{Mathematics of Gravitation. Part II. Gravitational
  Wave Detection}}, edited by
  \bibinfo{editor}{\bibfnamefont{A.}~\bibnamefont{Kr{\'o}lak}}
  (\bibinfo{publisher}{Polish Academy of Sciences, Institute of Mathematics,
  Warsaw}, \bibinfo{year}{1997}), pp. \bibinfo{pages}{71--74}.

\bibitem[{\citenamefont{Damour et~al.}(2000)\citenamefont{Damour, Jaranowski,
  and Sch{\"a}fer}}]{Damour-Jaranowski-Schaefer:2000b}
\bibinfo{author}{\bibfnamefont{T.}~\bibnamefont{Damour}},
  \bibinfo{author}{\bibfnamefont{P.}~\bibnamefont{Jaranowski}},
  \bibnamefont{and}
  \bibinfo{author}{\bibfnamefont{G.}~\bibnamefont{Sch{\"a}fer}},
  \bibinfo{journal}{Phys. Rev. D} \textbf{\bibinfo{volume}{62}},
  \bibinfo{pages}{084011} (\bibinfo{year}{2000}).

\bibitem[{\citenamefont{Monroe}(1976)}]{Monroe:1976}
\bibinfo{author}{\bibfnamefont{D.~K.} \bibnamefont{Monroe}}, Ph.D. thesis,
  \bibinfo{school}{University of North Carolina} (\bibinfo{year}{1976}).

\bibitem[{\citenamefont{Garat and Price}(2000)}]{Garat-Price:2000}
\bibinfo{author}{\bibfnamefont{A.}~\bibnamefont{Garat}} \bibnamefont{and}
  \bibinfo{author}{\bibfnamefont{R.~H.} \bibnamefont{Price}},
  \bibinfo{journal}{Phys. Rev. D} \textbf{\bibinfo{volume}{61}},
  \bibinfo{pages}{124011} (\bibinfo{year}{2000}).

\bibitem[{\citenamefont{Pfeiffer et~al.}(2000)\citenamefont{Pfeiffer,
  Teukolsky, and Cook}}]{Pfeiffer-Teukolsky-Cook:2000}
\bibinfo{author}{\bibfnamefont{H.~P.} \bibnamefont{Pfeiffer}},
  \bibinfo{author}{\bibfnamefont{S.~A.} \bibnamefont{Teukolsky}},
  \bibnamefont{and} \bibinfo{author}{\bibfnamefont{G.~B.} \bibnamefont{Cook}},
  \bibinfo{journal}{Phys. Rev. D} \textbf{\bibinfo{volume}{62}},
  \bibinfo{pages}{104018} (\bibinfo{year}{2000}).

\bibitem[{\citenamefont{Lousto and Price}(1998)}]{Lousto-Price:1998}
\bibinfo{author}{\bibfnamefont{C.~O.} \bibnamefont{Lousto}} \bibnamefont{and}
  \bibinfo{author}{\bibfnamefont{R.~H.} \bibnamefont{Price}},
  \bibinfo{journal}{Phys. Rev. D} \textbf{\bibinfo{volume}{57}},
  \bibinfo{pages}{1073} (\bibinfo{year}{1998}).

\bibitem[{\citenamefont{Matzner et~al.}(1999)\citenamefont{Matzner, Huq, and
  Shoemaker}}]{Matzner-Huq-Shoemaker:1999}
\bibinfo{author}{\bibfnamefont{R.~A.} \bibnamefont{Matzner}},
  \bibinfo{author}{\bibfnamefont{M.~F.} \bibnamefont{Huq}}, \bibnamefont{and}
  \bibinfo{author}{\bibfnamefont{D.}~\bibnamefont{Shoemaker}},
  \bibinfo{journal}{Phys. Rev. D} \textbf{\bibinfo{volume}{59}},
  \bibinfo{pages}{024015} (\bibinfo{year}{1999}).

\bibitem[{\citenamefont{Marronetti and
  Matzner}(2000)}]{Marronetti-Matzner:2000}
\bibinfo{author}{\bibfnamefont{P.}~\bibnamefont{Marronetti}} \bibnamefont{and}
  \bibinfo{author}{\bibfnamefont{R.~A.} \bibnamefont{Matzner}},
  \bibinfo{journal}{Phys. Rev. Lett.} \textbf{\bibinfo{volume}{85}},
  \bibinfo{pages}{5500} (\bibinfo{year}{2000}).

\bibitem[{\citenamefont{Gourgoulhon et~al.}(2002)\citenamefont{Gourgoulhon,
  Grandcl{\'e}ment, and Bonazzola}}]{Gourgoulhon-Grandclement-Bonazzola:2001a}
\bibinfo{author}{\bibfnamefont{E.}~\bibnamefont{Gourgoulhon}},
  \bibinfo{author}{\bibfnamefont{P.}~\bibnamefont{Grandcl{\'e}ment}},
  \bibnamefont{and}
  \bibinfo{author}{\bibfnamefont{S.}~\bibnamefont{Bonazzola}},
  \bibinfo{journal}{Phys. Rev. D} \textbf{\bibinfo{volume}{65}},
  \bibinfo{pages}{044020} (\bibinfo{year}{2002}).

\bibitem[{\citenamefont{Grandcl{\'e}ment
  et~al.}(2002)\citenamefont{Grandcl{\'e}ment, Gourgoulhon, and
  Bonazzola}}]{Grandclement-Gourgoulhon-Bonazzola:2001b}
\bibinfo{author}{\bibfnamefont{P.}~\bibnamefont{Grandcl{\'e}ment}},
  \bibinfo{author}{\bibfnamefont{E.}~\bibnamefont{Gourgoulhon}},
  \bibnamefont{and}
  \bibinfo{author}{\bibfnamefont{S.}~\bibnamefont{Bonazzola}},
  \bibinfo{journal}{Phys. Rev. D} \textbf{\bibinfo{volume}{65}},
  \bibinfo{pages}{044021} (\bibinfo{year}{2002}).

\bibitem[{\citenamefont{Bonazzola et~al.}(1993)\citenamefont{Bonazzola,
  Gourgoulhon, Salgado, and Marck}}]{Bonazzola-Gourgoulhon-etal:1993}
\bibinfo{author}{\bibfnamefont{S.}~\bibnamefont{Bonazzola}},
  \bibinfo{author}{\bibfnamefont{E.}~\bibnamefont{Gourgoulhon}},
  \bibinfo{author}{\bibfnamefont{M.}~\bibnamefont{Salgado}}, \bibnamefont{and}
  \bibinfo{author}{\bibfnamefont{J.-A.} \bibnamefont{Marck}},
  \bibinfo{journal}{Astron. \& Astrophys.} \textbf{\bibinfo{volume}{278}},
  \bibinfo{pages}{421} (\bibinfo{year}{1993}).

\bibitem[{\citenamefont{Bonazzola et~al.}(1998)\citenamefont{Bonazzola,
  Gourgoulhon, and Marck}}]{Bonazzola-Gourgoulhon-Marck:1998}
\bibinfo{author}{\bibfnamefont{S.}~\bibnamefont{Bonazzola}},
  \bibinfo{author}{\bibfnamefont{E.}~\bibnamefont{Gourgoulhon}},
  \bibnamefont{and} \bibinfo{author}{\bibfnamefont{J.-A.} \bibnamefont{Marck}},
  \bibinfo{journal}{Phys. Rev. D} \textbf{\bibinfo{volume}{58}},
  \bibinfo{pages}{104020} (\bibinfo{year}{1998}).

\bibitem[{\citenamefont{Kidder et~al.}(2000)\citenamefont{Kidder, Scheel,
  Teukolsky, Carlson, and Cook}}]{Kidder-Scheel-etal:2000}
\bibinfo{author}{\bibfnamefont{L.~E.} \bibnamefont{Kidder}},
  \bibinfo{author}{\bibfnamefont{M.~A.} \bibnamefont{Scheel}},
  \bibinfo{author}{\bibfnamefont{S.~A.} \bibnamefont{Teukolsky}},
  \bibinfo{author}{\bibfnamefont{E.~D.} \bibnamefont{Carlson}},
  \bibnamefont{and} \bibinfo{author}{\bibfnamefont{G.~B.} \bibnamefont{Cook}},
  \bibinfo{journal}{Phys. Rev. D} \textbf{\bibinfo{volume}{62}},
  \bibinfo{pages}{084032} (\bibinfo{year}{2000}).

\bibitem[{\citenamefont{Gourgoulhon et~al.}(2001)\citenamefont{Gourgoulhon,
  Grandcl{\'e}ment, Taniguchi, Marck, and
  Bonazzola}}]{Gourgoulhon-Grandclement-etal:2001}
\bibinfo{author}{\bibfnamefont{E.}~\bibnamefont{Gourgoulhon}},
  \bibinfo{author}{\bibfnamefont{P.}~\bibnamefont{Grandcl{\'e}ment}},
  \bibinfo{author}{\bibfnamefont{K.}~\bibnamefont{Taniguchi}},
  \bibinfo{author}{\bibfnamefont{J.-A.} \bibnamefont{Marck}}, \bibnamefont{and}
  \bibinfo{author}{\bibfnamefont{S.}~\bibnamefont{Bonazzola}},
  \bibinfo{journal}{Phys. Rev. D} \textbf{\bibinfo{volume}{63}},
  \bibinfo{pages}{064029} (\bibinfo{year}{2001}).

\bibitem[{\citenamefont{Kidder et~al.}(2001)\citenamefont{Kidder, Scheel, and
  Teukolsky}}]{Kidder-Scheel-Teukolsky:2001}
\bibinfo{author}{\bibfnamefont{L.~E.} \bibnamefont{Kidder}},
  \bibinfo{author}{\bibfnamefont{M.~A.} \bibnamefont{Scheel}},
  \bibnamefont{and} \bibinfo{author}{\bibfnamefont{S.~A.}
  \bibnamefont{Teukolsky}}, \bibinfo{journal}{Phys. Rev. D}
  \textbf{\bibinfo{volume}{64}}, \bibinfo{pages}{064017}
  (\bibinfo{year}{2001}).

\bibitem[{\citenamefont{Ansorg et~al.}(2002)\citenamefont{Ansorg,
  Kleinw{\"a}chter, and Meinel}}]{Ansorg-Kleinwaechter-Meinel:2002}
\bibinfo{author}{\bibfnamefont{M.}~\bibnamefont{Ansorg}},
  \bibinfo{author}{\bibfnamefont{A.}~\bibnamefont{Kleinw{\"a}chter}},
  \bibnamefont{and} \bibinfo{author}{\bibfnamefont{R.}~\bibnamefont{Meinel}},
  \bibinfo{journal}{Astron. \& Astrophys.} \textbf{\bibinfo{volume}{381}},
  \bibinfo{pages}{L49} (\bibinfo{year}{2002}).

\bibitem[{\citenamefont{Pfeiffer et~al.}(2002)\citenamefont{Pfeiffer, Kidder,
  Scheel, and Teukolsky}}]{Pfeiffer-Kidder-etal:2002}
\bibinfo{author}{\bibfnamefont{H.~P.} \bibnamefont{Pfeiffer}},
  \bibinfo{author}{\bibfnamefont{L.~E.} \bibnamefont{Kidder}},
  \bibinfo{author}{\bibfnamefont{M.~A.} \bibnamefont{Scheel}},
  \bibnamefont{and} \bibinfo{author}{\bibfnamefont{S.~A.}
  \bibnamefont{Teukolsky}} (\bibinfo{year}{2002}),
  \bibinfo{note}{gr-qc/0202096}.

\bibitem[{\citenamefont{Cook}(2000)}]{Cook:2000}
\bibinfo{author}{\bibfnamefont{G.~B.} \bibnamefont{Cook}},
  \bibinfo{journal}{Living Rev. Relativity} \textbf{\bibinfo{volume}{3}}
  (\bibinfo{year}{2000}), \bibinfo{note}{[Online Article]: cited on Aug 11,
  2001},
  \urlprefix\url{http://www.livingreviews.org/Articles/Volume3/2000-5cook/}.

\bibitem[{\citenamefont{Misner}(1963)}]{Misner:1963}
\bibinfo{author}{\bibfnamefont{C.~W.} \bibnamefont{Misner}},
  \bibinfo{journal}{Annals of Physics} \textbf{\bibinfo{volume}{24}},
  \bibinfo{pages}{102} (\bibinfo{year}{1963}).

\bibitem[{\citenamefont{Cook}(1991)}]{Cook:1991}
\bibinfo{author}{\bibfnamefont{G.~B.} \bibnamefont{Cook}},
  \bibinfo{journal}{Phys. Rev. D} \textbf{\bibinfo{volume}{44}},
  \bibinfo{pages}{2983} (\bibinfo{year}{1991}).

\bibitem[{\citenamefont{York{,}~Jr.}(2002)}]{York:2002}
\bibinfo{author}{\bibfnamefont{J.~W.} \bibnamefont{York{,}~Jr.}}
  (\bibinfo{year}{2002}), \bibinfo{note}{in preparation}.

\bibitem[{\citenamefont{Baumgarte et~al.}(1996)\citenamefont{Baumgarte, Cook,
  Scheel, Shapiro, and Teukolsky}}]{Baumgarte-Cook-etal:1996}
\bibinfo{author}{\bibfnamefont{T.~W.} \bibnamefont{Baumgarte}},
  \bibinfo{author}{\bibfnamefont{G.~B.} \bibnamefont{Cook}},
  \bibinfo{author}{\bibfnamefont{M.~A.} \bibnamefont{Scheel}},
  \bibinfo{author}{\bibfnamefont{S.~L.} \bibnamefont{Shapiro}},
  \bibnamefont{and} \bibinfo{author}{\bibfnamefont{S.~A.}
  \bibnamefont{Teukolsky}}, \bibinfo{journal}{Phys. Rev. D}
  \textbf{\bibinfo{volume}{54}}, \bibinfo{pages}{4849} (\bibinfo{year}{1996}).

\bibitem[{\citenamefont{Cook}(1994)}]{Cook:1994}
\bibinfo{author}{\bibfnamefont{G.~B.} \bibnamefont{Cook}},
  \bibinfo{journal}{Phys. Rev. D} \textbf{\bibinfo{volume}{50}},
  \bibinfo{pages}{5025} (\bibinfo{year}{1994}).

\bibitem[{\citenamefont{Brandt et~al.}(2000)\citenamefont{Brandt, Corell,
  G\'{o}mez, Huq, Laguna, Lehner, Marronetti, Matzner, Neilsen, Pullin
  et~al.}}]{Brandt-Corell-etal:2000}
\bibinfo{author}{\bibfnamefont{S.}~\bibnamefont{Brandt}},
  \bibinfo{author}{\bibfnamefont{R.}~\bibnamefont{Corell}},
  \bibinfo{author}{\bibfnamefont{R.}~\bibnamefont{G\'{o}mez}},
  \bibinfo{author}{\bibfnamefont{M.}~\bibnamefont{Huq}},
  \bibinfo{author}{\bibfnamefont{P.}~\bibnamefont{Laguna}},
  \bibinfo{author}{\bibfnamefont{L.}~\bibnamefont{Lehner}},
  \bibinfo{author}{\bibfnamefont{P.}~\bibnamefont{Marronetti}},
  \bibinfo{author}{\bibfnamefont{R.~A.} \bibnamefont{Matzner}},
  \bibinfo{author}{\bibfnamefont{D.}~\bibnamefont{Neilsen}},
  \bibinfo{author}{\bibfnamefont{J.}~\bibnamefont{Pullin}},
  \bibnamefont{et~al.}, \bibinfo{journal}{Phys. Rev. Lett.}
  \textbf{\bibinfo{volume}{85}}, \bibinfo{pages}{5496} (\bibinfo{year}{2000}).

\bibitem[{\citenamefont{Damour et~al.}(2002)\citenamefont{Damour, Gourgoulhon,
  and Grandcl{\'e}ment}}]{Damour-Gourgoulhon-Grandclement:grqc0204011}
\bibinfo{author}{\bibfnamefont{T.}~\bibnamefont{Damour}},
  \bibinfo{author}{\bibfnamefont{E.}~\bibnamefont{Gourgoulhon}},
  \bibnamefont{and}
  \bibinfo{author}{\bibfnamefont{P.}~\bibnamefont{Grandcl{\'e}ment}},
  \bibinfo{journal}{gr-qc/0204011}  (\bibinfo{year}{2002}).

\bibitem[{\citenamefont{Cook}(2002)}]{Cook:2001}
\bibinfo{author}{\bibfnamefont{G.~B.} \bibnamefont{Cook}},
  \bibinfo{journal}{Phys. Rev. D} \textbf{\bibinfo{volume}{65}},
  \bibinfo{pages}{084003} (\bibinfo{year}{2002}).

\end{thebibliography}

\end{document}